\documentclass[11pt]{article}

\usepackage{mathpazo}
\usepackage{fullpage}
\usepackage{natbib}
\usepackage{booktabs} 
\usepackage[ruled]{algorithm2e} 
\usepackage{enumitem}
\usepackage{microtype}
\usepackage{graphicx}
\usepackage{subcaption}
\usepackage{sidecap, caption}
\usepackage{booktabs} 
\usepackage{array}
\usepackage{wrapfig}
\usepackage{sidecap}
\usepackage{float}
\usepackage{hyperref}
\usepackage{url}
\usepackage{amsmath,amssymb,amsfonts}

\SetAlFnt{\small}
\SetAlCapFnt{\small}
\SetAlCapNameFnt{\small}
\SetAlCapHSkip{0pt}
\IncMargin{-\parindent}

\setcitestyle{authoryear}

\title{Diffusion Based Auctions for Multi-Winner Banner Ads\footnote{Appeared in the proceedings of the 27th ACM Conference on Economics and Computation}}

\author{Lilian Sun\\Harvard \and Henry Huang\\Harvard \and Fucheng Warren Zhu\\Harvard \and Giannis Daras\\MIT
\and Constantinos Daskalakis\footnote{Supported by a Simons Investigator Award, a Simons Collaboration on Algorithmic Fairness, ONR MURI grant N00014-25-1-2116 and ONR grant N00014-25-1-2296.}\\MIT}

\begin{document}


\maketitle

\begin{abstract}
We propose a new framework for banner ad auctions, leveraging diffusion models to allow for multiple bidders to co-win a single banner ad opportunity, and in particular influence the ad shown on the banner depending on the strength of their bids, enabling a far richer outcome space compared to standard banner ad auctions, where there is a single winner who shows their ad. In our framework, the outcomes are images (what ad to show) and each bidder expresses their preferences over outcomes by declaring their value per unit of alignment between the image chosen by the mechanism and a textual description of their desired image. Our proposed mechanism combines the advertisers’ bids and textual descriptions into a score that drives the reverse process of a diffusion model. This is used to sample a set of candidate images from which the welfare maximizing one is eventually chosen. Extensive experimentation evaluates our proposed mechanism around the following axes: (1) efficiency improvements of our allocation rule relative to the single-winner allocation rule; (2) monotonicity of our allocation rule; and (3) approximate incentive compatibility when combining our allocation rule with Myerson lemma-based or Clarke rule-based payments. Improvements along all these axes are shown with the number of Monte Carlo samples. Meanwhile, these benefits are attained while preserving the image quality.

\end{abstract}

\newpage
\vspace{1cm}
\setcounter{tocdepth}{1} 
\tableofcontents
\newpage

\section{Introduction}

Online advertising auctions, including sponsored search and banner ad auctions, are a cornerstone of the digital economy. A fundamental limitation of these auctions, however, is that they make \emph{discrete} allocations, in that every  ad opportunity is allocated to a single advertiser. However, this might leave economic efficiency on the table. It stands to reason that augmenting the allocation space to allow multiple advertisers to co-win an advertising opportunity and co-influence the shown ad according to their combined preferences could substantially improve efficiency in many contexts. The challenge is that, in order to unlock this source of increased efficiency, the allocation rule of the mechanism needs to optimize over images to output one that is both natural and aggregates the advertisers' preferences according to the strength of their bids. The goal of this paper is to investigate how to do this by combining methods from mechanism design and diffusion models. 






Taking a step back, our work belongs to a growing line of literature investigating the interface between mechanism design and generative AI. While monetization of generative AI is certainly one motivation for pursuing this line of research, our motivation is different. We want to explore whether generative AI can be leveraged to overcome the limitation of optimizing over a discrete set of candidate allocations, encountered in many mechanism design settings involving digital goods. When a  generative model for this type of goods is available, the outcome of the auction can be taken to be much richer and the allocation of the auction can leverage the generative model to select from this richer set of alternatives. 

Consider banner ads, for example. In a standard banner ad auction, the allocation rule of the auction effectively selects one (or none) out of a finite set of banners corresponding to the different bidders, as there is at most one winner of the auction. Instead, we want to think of the outcome space as continuous, allowing all possible images to be placed on the banner! By allowing a larger set of outcomes, we can combine the preferences of multiple bidders in the chosen allocation, and increase the efficiency, or other economic property of the auction.

Analogous opportunities arise for other data modalities.
Consider, e.g., asking ``\emph{What should I do in Boston?}'' to a Large Language Model (LLM). The LLM's response could be:
\begin{quote}
\small
Boston is a great city for Italian food$\dots$ Try \texttt{\{<Italian Restaurant Ad 1>\}} in the North End, then stop by \texttt{\{<Local Shop Ad 2>\}} for souvenirs.
\end{quote}
By allowing multiple bidders to influence the text generation process, we can allow multiple advertisers to win exposure from the answer to a single query. Indeed, recent works by~\citet{duetting2024mechanismdesignlargelanguage,hajiaghayi2024adauctionsllmsretrieval,feizi2024onlineadvertisementsllmsopportunities,dubey2024auctionsllmsummaries} make various different proposals for how to use mechanism design methods in auto-regressive generation to enable multiple advertisers to influence the sampled text according to their bid strength.

\smallskip Our goal in this paper is to study the visual domain and, in particular, to explore how {\em diffusion models}, as opposed to autoregressive ones, can be leveraged to enable mechanisms whose allocation space is the continuum of all possible images.
To fix ideas, consider the banner ad auction example discussed earlier. We want to design an auction whose goal is to select what content to show on a banner, when there are several advertisers interested to show their advertising content on the banner, and we want to allow any possible content to appear on the banner (subject perhaps to certain ground rules that prevent offensive or otherwise undesirable content). To approach this question we need to model how different bidders express their preferences, how the allocation rule of the auction aggregates these preferences, and how the bidders are charged. Our approach and contributions are as follows:

\smallskip \noindent $\bullet$ \textbf{A novel auction design challenge:} We introduce an auction design challenge where multiple bidders compete for the generation of an image. In our stylized model, each bidder $i$ has a private scalar type $m_i$ and a public caption (prompt) $c_i$. So the setting is {\em single-dimensional}. The bidder's value for an image $I$ that the auction can select is $m_i \cdot \alpha(I,c_i)$, i.e.~the alignment $\alpha(I,c_i) \in [0,1]$ between the image and their caption, multiplied by their private type. The alignment can be computed using a standard model that computes the alignment between images and captions, e.g.~CLIP, and it is fixed by the auctioneer. The bidders are asked to report their value $m_i$ per unit of alignment according to this model. The allocation space of the auction are all possible images, of some fixed resolution. The goal of the auctioneer is to maximize welfare. Perhaps the auctioneer has some basic desiderata for the image, e.g.~enforcing some content, avoiding certain outputs, etc. These are all summarized in a ``system caption'' $c_{\emptyset}$ and a dummy bidder is included in the computation of the welfare, whose value is $m_{\emptyset} \cdot \alpha(I,c_\emptyset)$. Besides the system caption there is also an implicit desideratum that the image be ``natural.''
    
    
    
\begin{figure}[h!]
    \centering
    \begin{subfigure}[b]{0.193\textwidth}
        \centering
        \includegraphics[width=\textwidth]{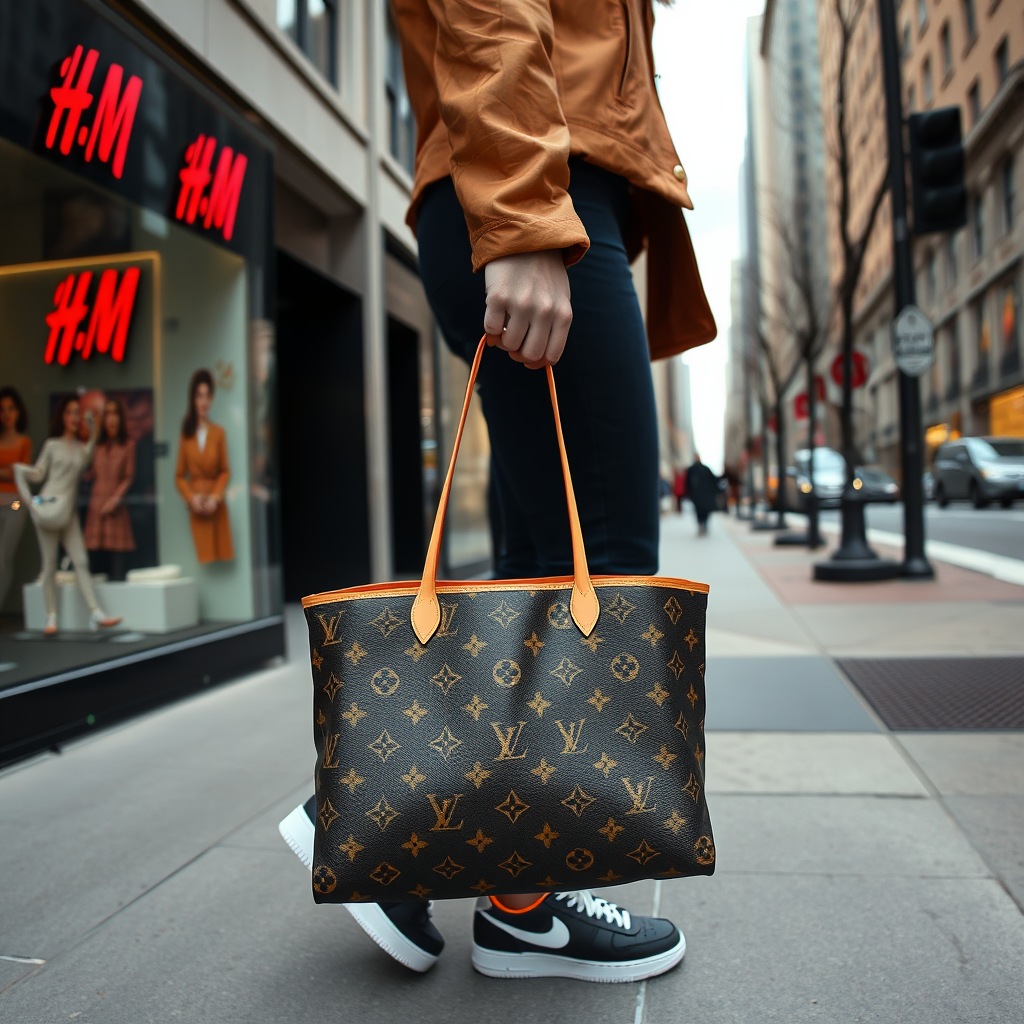}
        \caption{$(0.33, 0.33, 0.33)$}
        \label{fig:sub1_3}
    \end{subfigure}
    \begin{subfigure}[b]{0.193\textwidth}
        \centering
        \includegraphics[width=\textwidth]{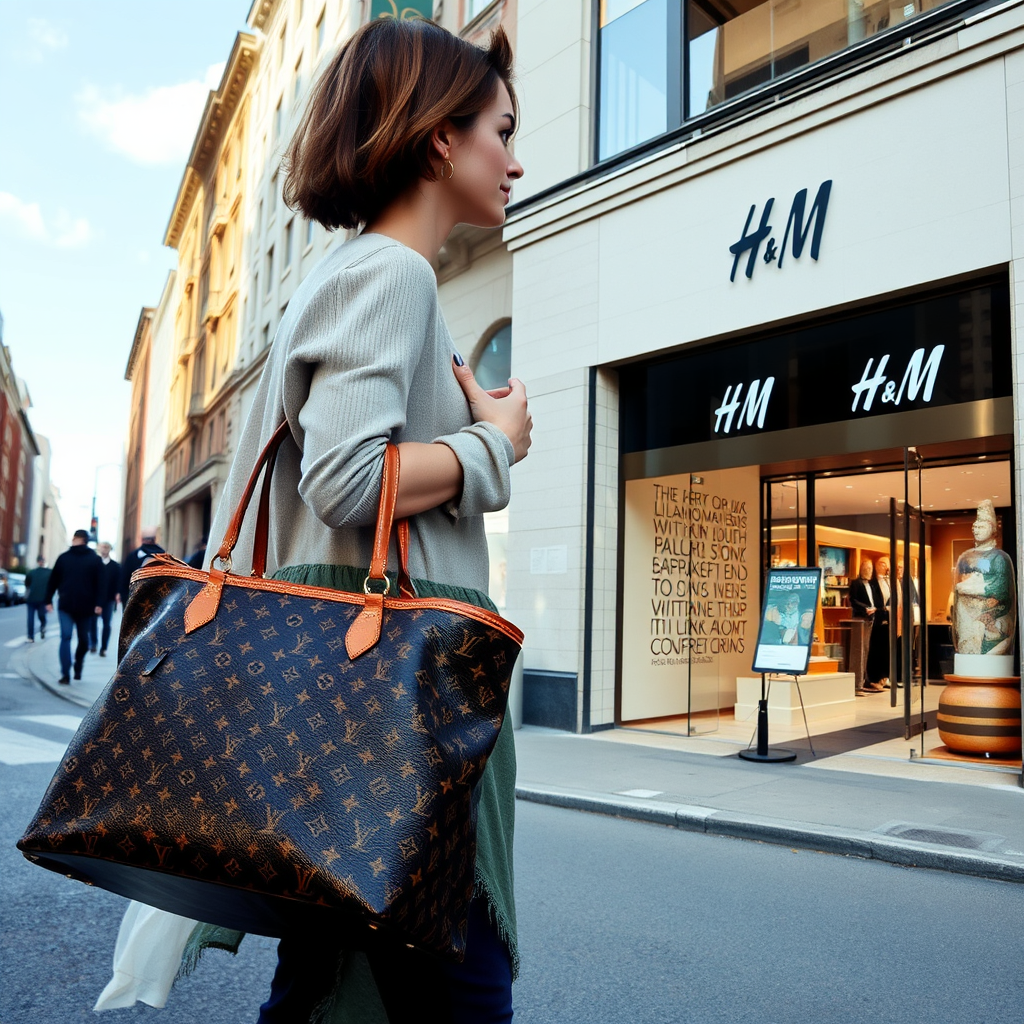}
        \caption{$(0.4, 0.4, 0.2)$}
        \label{fig:sub2_3}
    \end{subfigure}
    \begin{subfigure}[b]{0.193\textwidth}
        \centering
        \includegraphics[width=\textwidth]{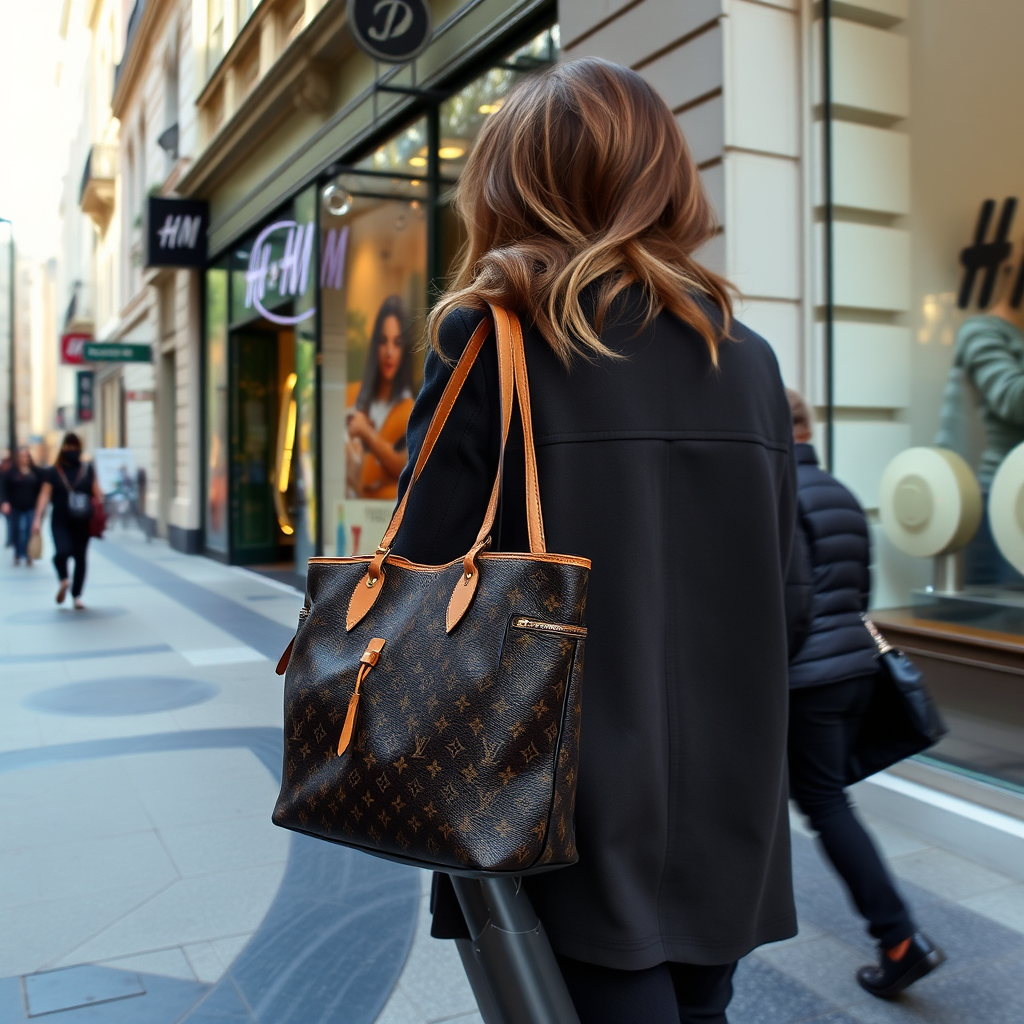}
        \caption{$(0.6, 0.3, 0.1)$}
        \label{fig:sub3_3}
    \end{subfigure}
    \begin{subfigure}[b]{0.193\textwidth}
            \centering
            \includegraphics[width=\textwidth]{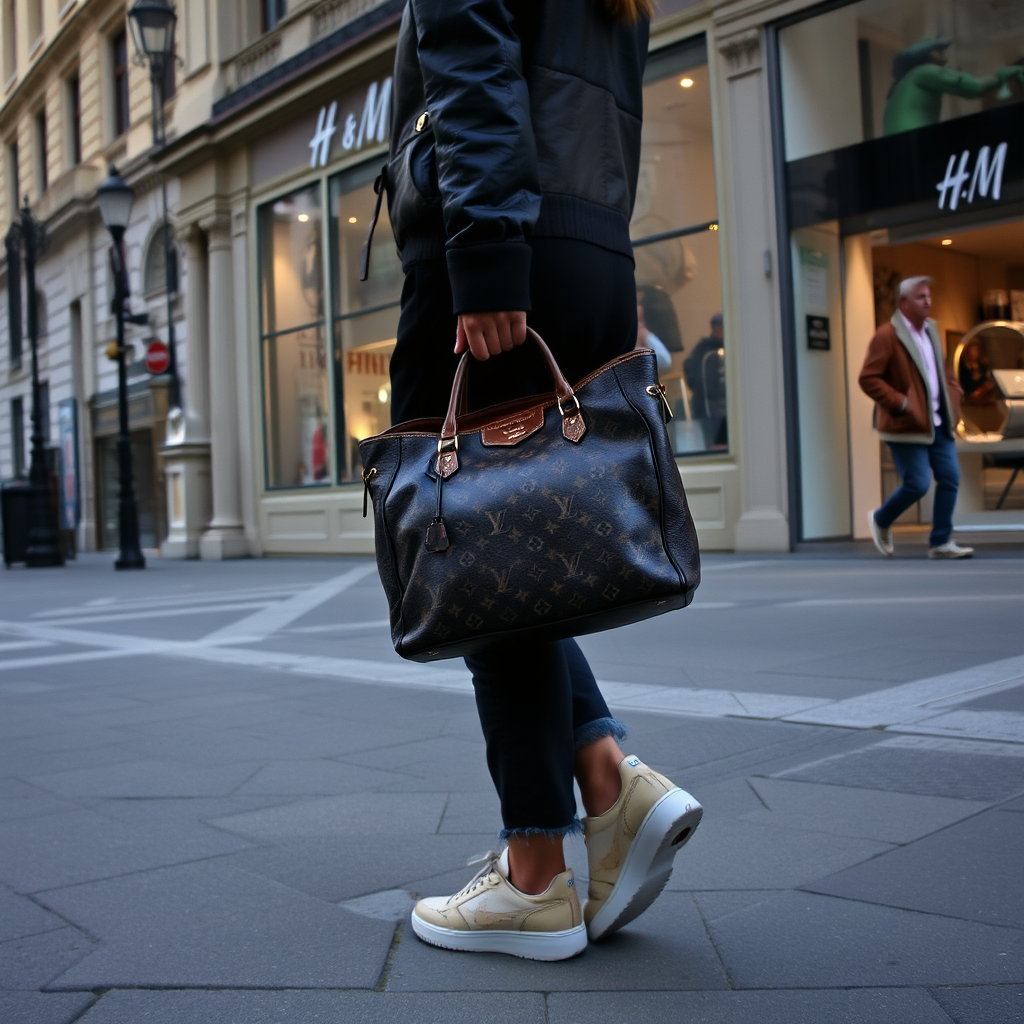}
            \caption{$(0.6, 0.2, 0.2)$}
            \label{fig:sub4_3}
        \end{subfigure}
    \begin{subfigure}[b]{0.193\textwidth}
        \centering
        \includegraphics[width=\textwidth]{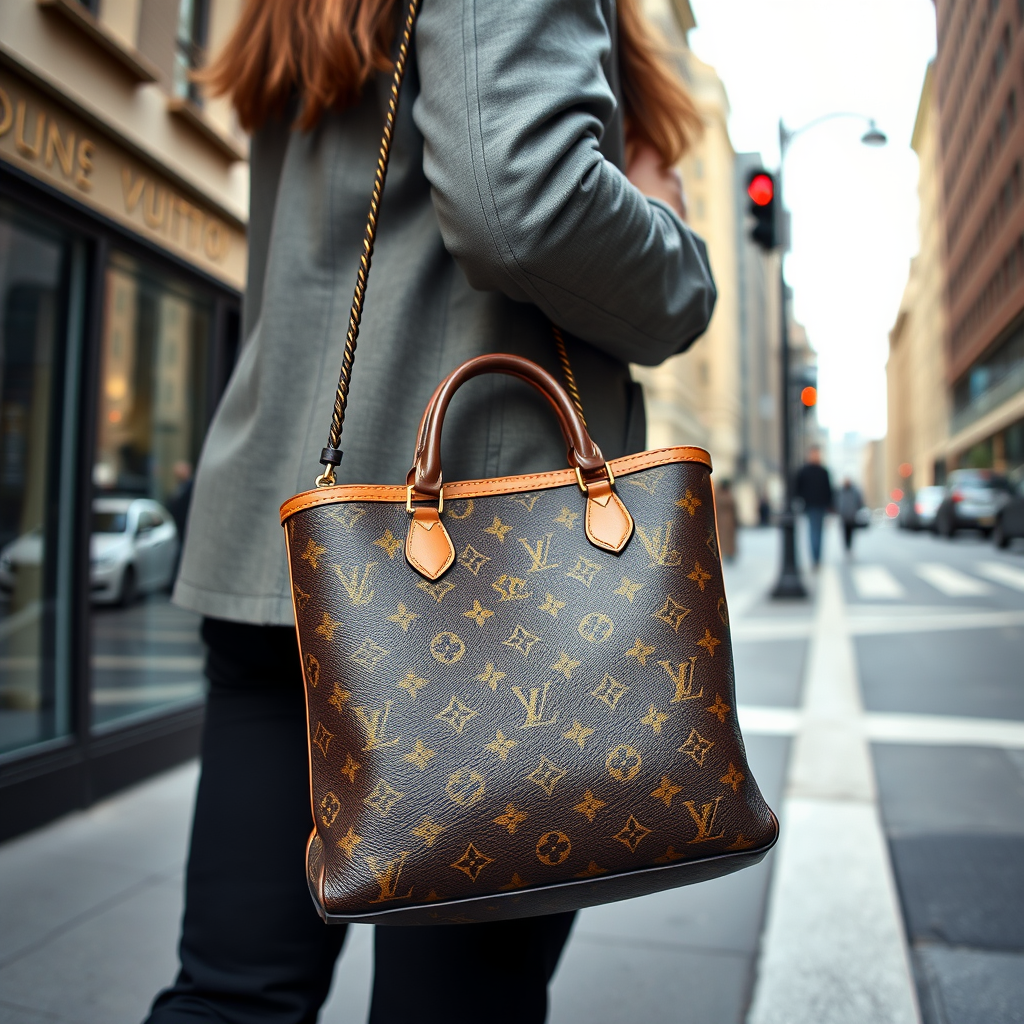}
        \caption{$(1, 0, 0)$}
        \label{fig:sub5_3}
    
    \end{subfigure}
    
    \caption*{Sampled images for $3$ agents with captions ``Louis Vuitton bag'', ``H\&M storefront'' and ``Nike shoes'' and (normalized) bids $b_1$, $b_2$ and $b_3$ respectively, for different choices of $(b_1,b_2,b_3)$.}
    \caption{Visual examples of our allocation rule. This scenario simulates an auction between three agents with captions ``Louis Vuitton bag'', ``H\&M storefront'' and ``Nike shoes'' bidding to influence the generation of an image with system caption ``A person walking down a city street.'' We denote by $b_1$, $b_2$ and $b_3$ the normalized bids of the agents, i.e.~the relative strengths of their bids. From left to right, the images show the visual output as the bid distribution shifts from balanced, where all agents are equally represented, to scenarios with one or two dominant agents, and finally to a single strong bidder case. This fine-grained, bid-dependent control is achieved by our mechanism aggregating agent preferences directly within the diffusion model's score function, creating a scalable auction for a continuous, visual outcome.}
    \label{fig:grid_images_3_agents}
\end{figure}

\smallskip \noindent $\bullet$ \textbf{An allocation rule:} We propose a scalable, diffusion model-based generative algorithm that can serve as the allocation rule of an auction in the afore-described setting. Through comprehensive empirical validation, we show that our algorithm generates images that are (i) visually appealing; (ii) aggregate the bidders' preferences according to their captions and the strengths of their bids; and in particular (iii) attain higher welfare compared to the single-winner alternative where the highest bidder wins and an image is generated according to their caption. Figure~\ref{fig:grid_images_3_agents} demonstrates (i) and (ii), showing the outputs of our algorithm for three bidders of different captions and varying bids. In Section~\ref{sec:efficiency}, we provide extensive experimental evidence for (iii).

  \noindent Our allocation rule is designed as follows. We make use of a conditional diffusion model,\footnote{Specifically, for all our experiments we use \texttt{FLUX.1-schnell}.} sampling a distribution, $p(x|c)$, over images $x$ conditioning on captions $c$. For each bidder's caption~$c_i$, this model provides us  access to a conditional score function $s_t(x|c_i):=\nabla_x \log p_t(x|c_i)$, where $t$ indexes diffusion time; in particular $p_t(x|c) = p(x|c) \ast {\mathcal N}(0,\sigma^2(t){\rm I})$ for some increasing noise schedule $\sigma^2(t)$, $t\in [0,T]$. Using ideas from classifier-free guidance~\cite{ho2022cfg}, we aggregate these bidder-specific conditional scores according to the relative strengths of their bids to obtain a composite score function, which we use to define a reverse diffusion process. We run this reverse diffusion process for a number of times, $k$, to sample multiple images, and output the image with the highest welfare. Our algorithm is scalable. Its generation time depends on the time taken to run the reverse diffusion process times the number of samples $k$ that we draw. Meanwhile, the number of samples influence the final welfare, as shown by our experiments. So there is a trade-off between generation time and welfare.
    
\begin{figure}[h!]
        \centering
        \includegraphics[width=0.8\linewidth]{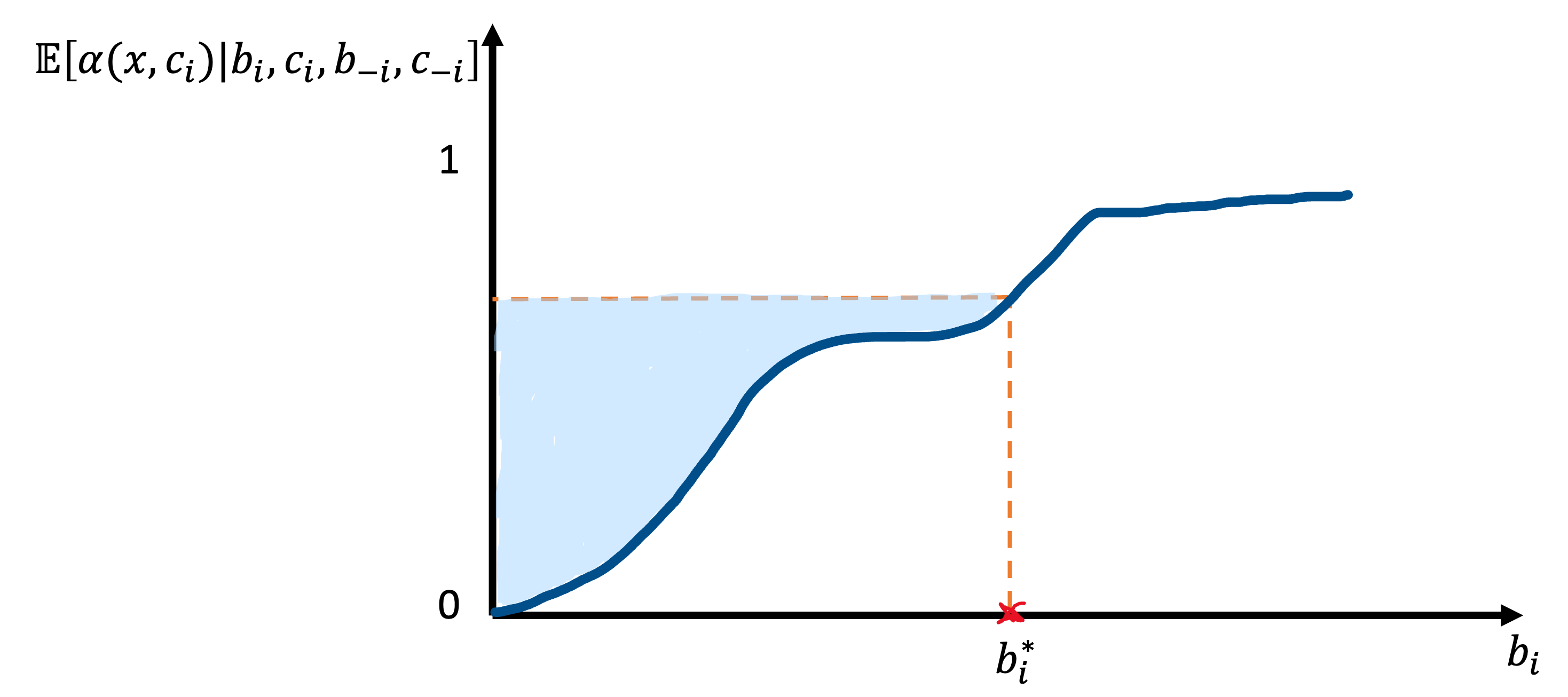}
        \caption{If the expected alignment between the generated image and bidder $i$'s caption were increasing in bidder $i$'s bid, we would be able to straightforwardly adapt the payment rule from Myerson's lemma to compute prices that induce  truthful reporting of types a dominant strategy.}
        \label{fig:myerson's lemma}
    \end{figure}

\smallskip \noindent $\bullet$ \textbf{Payment rules:} The well-studied empirical properties of classifier-free guidance~\cite{ho2022cfg} suggest that, as bidder $i$ increases their bid $b_i$ while the bids $b_{-i}$ of the other bidders remain fixed, the image sampled from the reverse diffusion process defined by our composite score is influenced more and more by bidder $i$'s caption $c_i$. So the expected alignment $\alpha(x,c_i)$ between the generated image, $x$, and bidder $i$'s caption should be approximately increasing in bidder $i$'s bid~$b_i$, everything else held constant. Suppose that it was indeed increasing;  in Figure~\ref{fig:bid_monotonicity} we provide evidence that, at least approximately, this is the case. Then, because the alignment score lies in $[0,1]$, and the bidder's value is their private type times the alignment $\alpha(x,c_i)$, we can straightforwardly adapt the payment rule from Myerson's lemma to  compute prices that make truthful reporting of types a dominant strategy. Figure~\ref{fig:myerson's lemma} shows that this adaptation is straightforward.
    In practice, the expected alignment might only be approximately increasing in the bid, because the diffusion model and the alignment model might be misaligned (no pun intended) with each other. So, if we stick to the payment rule that is the area to the left of the curve, we will have approximate incentive compatibility. There are two additional issues we need to consider: (1) We only have sample access to $\alpha(x,c_i)$ for a given set of bids and captions. This is not a big issue. We can apply Myerson's lemma on a sampled curve (whose randomness is with respect to the randomness in running the diffusion process $k$ times and for various bids by the bidder) and we would get the desired price, in expectation. (2) We do not have an analytical description of the expected curve or a sampled curve to apply Myerson's lemma.  So we will need to do numerical integration to estimate the price, which will introduce error in incentive compatibility. The errors resulting from the approximate monotonicity and the numerical integration should be small, as illustrated in Figure~\ref{fig:myerson_regret}, and the latter is also controllable with computational effort. 
    
    
    \smallskip \noindent The main disadvantage of the afore-described approach is that it takes time to perform numerical integration, and this would make the price computation slow. As a heuristic alternative, we propose using prices inspired by the VCG payment rule. For each bidder $i$, we run our allocation rule without this bidder's bid or caption to sample an image. We then charge this bidder the difference between the welfare of the other bidders  under this image and under the image sampled with this bidder included. This is an approximation to the VCG allocation rule, which exactly optimizes  over images. Instead, we optimize using Monte Carlo sampling. Due to this approximation, 
    this will not be guaranteeing incentive compatibility. We perform extensive experimentation to evaluate how approximately truthful the resulting mechanism is. See Section~\ref{sec:truthfulness}. As shown in our experiments, the incentive to not be truthful decreases with the number $k$ of Monte Carlo samples in the allocation rule, and is relatively controlled. The additional benefit of a VCG-style payment is that it is also robust to manipulation of captions, if we are in a setting where bidders can manipulate both their value per unit of alignment of the chosen image with their caption {\em and} their caption. 

 \smallskip \noindent  $\bullet$ \textbf{Extensive empirical validation:} As mentioned earlier, we perform extensive experimentation to evaluate the performance of our proposed allocation rule and mechanism. In particular, we experiment with a wide range of prompts capturing settings where the bidders have prompts that are non-competing (e.g.~``Nike shoes'' and ``Gucci bag'') or competing (e.g.~``Starbucks coffee'' and ``Dunkin coffee''). We consider settings where bidders' bids are comparable in strength or there is  variation in their strengths. We experiment with settings from two up to twenty bidders. Finally, we experiment with the number of Monte Carlo samples used in our allocation rule.  We demonstrate through extensive multi-agent experiments on a diverse dataset that our mechanism achieves key economic properties: welfare improvement over single-winner baselines, approximate incentive compatibility, and bid monotonicity, all without degrading image quality. 
 
 \medskip All in all, our work proposes a new challenge in mechanism design, and a productive combination of diffusion model and mechanism design methods to approach it, complete with extensive experimental evidence about the economic value of extending beyond single-winner mechanisms. We believe that we make an important first step to a very rich direction.

\section{Related Work}

We review the most relevant strands of related literature, focusing on recent work at the interface of mechanism design and Large Language Models, and guiding diffusion models. 


\subsection{Mechanism Design and Generative AI}

Recent research has begun to explore the interface of mechanism design and Large Language Models, largely motivated by the monetization of LLMs. A central challenge is how to model and how to aggregate the preferences of agents over generated content. To this end, \citet{duetting2024mechanismdesignlargelanguage} introduce token auctions, wherein the bidders are competing at the token level, competing to influence the distribution of the next generated token. Others have focused on contextually relevant ad placement within LLM outputs using retrieval-augmented generation~\citep{hajiaghayi2024adauctionsllmsretrieval}; auctions for placement within ad summaries \citep{dubey2024auctionsllmsummaries}; and auctions where the bidders are bidding for modifications to a neutrally generated text~\citep{feizi2024onlineadvertisementsllmsopportunities}. While these works have provided a starting point for studying mechanism design in generative applications involving text, they have left open the question of how to do so in generative applications involving other data modalities and other types of generative models, beyond LLMs. This is the challenge that we set out to study in this work, focusing on image generation and diffusion models. Moreover, our main motivation is to enhance the attainable welfare in auction settings, such as banner ad auctions, that are already widely used in practice, by enabling an increased allocation space, versus introducing advertising to generative models which is the motivation of some of the afore-described works.

\subsection{Controllable Image Generation with Diffusion Models}

To realize a generative auction in the visual domain, we build upon the recent advances in diffusion models, which have achieved state-of-the-art results in synthesizing diverse, high-fidelity images~\citep{ho2020ddpm, song2019generative, rombach2022highresolutionimagesynthesislatent, dhariwal2021diffusionmodelsbeatgans}. These models learn how to generate data by learning how to reverse a gradual noising process, which allows them to progressively remove noise from a random sample until a clean image sample is produced~\citep{sohldickstein2015deepunsupervisedlearningusing, ho2020ddpm, songSohlDickstein2021score}. Recent advances, such as Rectified Flow, have further improved generation speed and quality \citep{liu2022flowstraightfastlearning, esser2024scalingrectifiedflowtransformers}.

An important feature of this technology for our purposes is its controllability. Generation can be steered toward desired attributes using guidance techniques. Classifier-free guidance (CFG) has become a prominent method, as it allows for strong conditional control without a separate classifier model \citep{ho2022cfg}. Its effectiveness in high-fidelity caption-to-image synthesis has been well-demonstrated \citep{saharia2022photorealistictexttoimagediffusionmodels, nichol2022glidephotorealisticimagegeneration}. Our work leverages this technical capability not just for creative control, but as a primitive upon which we build an allocation rule that integrates multiple agents' preferences.


\section{Methodology}

Recall that our goal in this paper is to propose an auction that takes as input the preferences of bidders over images and generates a single image that aggregates these preferences in a welfare-maximizing way. We will propose a model specifying how the agents may express their preferences over images to the auction, an algorithm that aggregates these preferences to sample an image with high social welfare (the allocation rule), and an algorithm computing the agents' payments (the payment rule). Our goals from the proposed auction are (a) approximate monotonicity of the allocation rule; (b) approximate incentive compatibility of the auction; and (c) welfare-improvement compared to single-winner baselines, i.e. auctions that select the highest bidder as the winner and generate an image according to this bidder's preferences. We start with some preliminaries from auction theory and diffusion models, before presenting our proposed design.

\subsection{Preliminaries}
\label{sec:preliminaries}

\subsubsection{Auctions and Social Welfare}
Our goal is to maximize {\em social welfare}. For a set of agents $N$ and a chosen outcome $I$, which in our case is a {\em natural image}, the social welfare is 
\begin{equation}
W(I) = \sum_{i \in N} v_i(I),
\end{equation}
where $v_i(I)$ is the value that agent $i$ derives from image $I$. An auction takes as input some description of the $v_i$'s and computes an outcome $I$ as well as payments $\{p_i\}_{i \in N}$ for all the bidders. The {\bf allocation rule} of the auction computes $I$ and the {\bf payment, or price, rule} outputs the prices.  If agents could communicate their $v_i$'s exactly and we were able to optimize over natural images, we could use the Vickrey-Clarke-Groves (VCG) mechanism to optimize social welfare in a truthful fashion. Unfortunately, optimizing $W(I)$ over $I$ is highly non-trivial. One reason is that it is not mathematically easy to define what makes $I$ a natural image. So it is unclear what is the space of feasible allocations. Another challenge is that it is not clear how bidders can specify their preferences over natural images.

\subsubsection{Score-Based Diffusion and Guidance}
Denoising diffusion models (see references above) learn to sample a target distribution $p(x)$ over images to which sample access is provided, by learning how to reverse a diffusion process that progressively adds noise to samples from  $p$ according to a noise schedule $\sigma^2(t)$. This is often achieved by learning the score functions, $s_t(x) = \nabla_{x} \log p_t(x)$, of the distributions, $p_t(x):=p(x)\ast {\mathcal N}(0,\sigma^2(t){\rm I})$, of noised samples. Knowledge of the scores suffices to run the reverse diffusion process, i.e.~starting with a noisy sample and progressively improving it until a clean sample is obtained. There are variations to the theme related to whether the noising process is variance exploding (as we presented it) or variance preserving or has a different type of drift, and various other generalizations of this theme. 

When caption-image data is available, one targets a model that can do conditional generation, i.e.~sample an image $x$  conditioning on a caption $c$. To gain this capability, the approach is to learn the scores, $s_t(x|c) = \nabla_{x} \log p_t(x|c)$,  of the noised conditionals, $p_t(x|c):=p(x|c)\ast {\mathcal N}(0,\sigma^2(t){\rm I})$.

 Once the conditional scores are learned, they can be used directly to do conditional sampling by reversing the diffusion process starting with a sample from $p(x|c)$, where $c$ is the caption that we want to condition on. A key technique that has been found to improve the generated samples is Classifier-Free Guidance (CFG) \citep{ho2022cfg}. CFG sharpens the generation by steering the unconditional score $s_t(x|\emptyset)$ towards the conditional score $s_t(x|c)$.  The guided score $\hat{s}_t$ is typically taken to be:
\begin{equation}
    \hat{s}_t(x|c) = (1-\omega) s_t(x|\emptyset) + \omega s_t(x|c) \label{eq:CFG}
\end{equation}
where $\omega$ denotes the guidance scale. Increasing $\omega$ amplifies the conditioning signal.

\subsection{Our Setting: Image Generation Auctions}
We now formalize the main components of our setting:
\begin{enumerate}[leftmargin=6mm, nosep]
    \item \textbf{Allocation Space ($\mathcal{I}$):} The goal of the auction is to output a {\em natural} image $I \in \mathcal{I}$, where $\mathcal{I}$ is the space of all ``natural images.'' Part of the challenge here is that the set of natural images is not well-defined mathematically, and we only have proxies for accessing this set, e.g.~by sampling a diffusion model trained to sample natural images. 
     \item \textbf{Agents ($N = \{1, \dots, n\}$):} Each agent $i$ wishes to influence the image output by the auction.
    \item \textbf{Agent Types:} Each agent $i$ has a {\em public} text prompt/caption $c_i$ describing their desired content and a {\em private} type $m_i \in \mathbb{R}_{\ge 0}$ which is a scalar. In particular, the setting is single-dimensional.\footnote{While this is the main setting we will study, it is well motivated to consider the multi-dimensional setting where the caption of the bidder is also private. We will briefly discuss this setting in this paper.}
    \item \textbf{Agent Valuations:} As a function of their public caption, $c_i$, and private type, $m_i$, bidder $i$'s valuation is $v_i(I) = m_i \cdot \alpha(I,c_i)$, where $\alpha(I,c_i) \in [0,1]$ is how well image $I$ aligns with caption $c_i$ according to some fixed model (announced to the bidders) that computes image-text alignment. We will be using the CLIP model throughout. In particular, a bidder's private type expresses their value per unit of alignment between the image and their caption.
    \item \textbf{Agent Bids:} Each bidder will communicate a bid $b_i \in \mathbb{R}_{\ge 0}$ to the auction.
    \item \textbf{Base/System Prompt ($c$):} A text prompt describing the base content of the image to be output, as well as guardrails protecting against offensive content. This could be empty. We could view the system  prompt as an extra bidder with valuation $v_0(I)=m_0 \cdot \alpha(I,c)$.

\end{enumerate}

\medskip \noindent In the context defined by the above ingredients, the goal of our mechanism is to implement an \textbf{Allocation Rule} that will generate a high-welfare image (Section~\ref{sec:allocation_rule}) and a \textbf{Payment Rule} that will charge the agents in a way that induces good incentive properties (Section~\ref{sec:payment_rule}). We propose such rules in the next few sections. 

\subsection{A Prelude: the VCG auction}
\label{sec:prelude}

A natural question to ask at this point is what it would mean to implement the Vickrey-Clarke-Groves mechanism. The allocation rule of this mechanism would be to choose an image as follows:
\begin{align}
    I^* \in \arg\max_{I \in \mathcal{I}}\sum_{i \in N \cup \{0\}} v_i(I), \label{eq: welfare optimal allocation}
\end{align}
using some tie-breaking rule, while the Clarke payment rule would charge bidder $i$ as follows:
\begin{align}
    p_i =  \max_{I \in \mathcal{I}}\sum_{j \neq i} v_j(I) - \sum_{j \neq i} v_j(I^*). \label{eq: Clarke payment rule}
\end{align}
The challenge is that the maximization problems showing up in the allocation and price rules above are over $\mathcal{I}$, the set of ``natural images.'' Unfortunately, this set does not have a mathematical specification which is amenable to exact optimization. We will instead leverage diffusion models to access this set, and approximately optimize over it via sampling. In the next section, we discuss how to build effective samplers. 

\subsection{Our Allocation Rule: From Bids and Captions to High-Welfare Images via Score Composition}
\label{sec:allocation_rule}

Our allocation rule approximately optimizes welfare over natural images, by leveraging a diffusion model, and in particular a text-to-image diffusion model. At the heart of our approach lies a score composition technique, inspired by classifier-free guidance. Our score composition technique aggregates bidders' bids and captions into a composite score that drives a reverse diffusion process, used to sample over high-welfare parts of $\mathcal{I}$.

We first describe our score composition technique for two agents as it is simpler. Instead of combining conditional scores with the unconditional score (as is done in classifier-free guidance, see Eq.~\eqref{eq:CFG}), we perform a bid-weighted interpolation between two conditional scores: the conditional score with the concatenation of the bidders' prompts and the conditional score under the highest bidder's prompt.  More precisely, suppose that our two agents submitted bids $b_1, b_2$, where $b_1 \ge b_2$ without loss of generality. We define bidder $1$'s normalized bid $b_1^{(1)} = \frac{b_1}{b_1+b_2}$ and, from this, bidder 1's weight $w^{(1)} = 2b_1^{(1)}-1$. We then define our composite score $s_t^{(1,2)}(x)$ as follows:
\begin{equation}
    s^{(1,2)}_t(x) = (1-w^{(1)})s_t(x|c,c_{1,2}) + w^{(1)} s_t(x|c,c_1)
\end{equation}
where $c,c_{1,2} = \{c,c_1, c_2\}$ is the concatenated prompt of the two bidders (and the base prompt if any), and $c,c_1=\{c,c_1\}$ is the concatenation of the base prompt and bidder 1's prompt. When the bids are equal, we have $b_1^{(1)}=0.5$ and $w^{(1)}=0$ so the composite score relies solely on the first term. When Agent 1 is fully dominant, i.e.~$b_1^{(1)}=1$, we have $w^{(1)}=1$, and the composite score collapses to the second term, i.e.~the dominant agent's score.

We generalize the score composition technique to $n$ agents recursively, with the full derivation in Appendix~\ref{app:generalized_score}. The score for $n$ agents, whose bids are w.l.o.g.~ordered as $b_1 \ge b_2 \ge \cdots \ge b_n$, is defined as:
\begin{equation}
s^{(1:n)}_t(x) = (1-w^{(1)}) \, s(x|c, c_{1:n}) + w^{(1)} \, s^{(1:n-1)}_t(x)
\end{equation}
where $s^{(1:n-1)}_t(x)$ is the composite score for the $n-1$ most dominant agents, and it is computed recursively.

Given our composite score, we are ready to describe the allocation rule of our mechanism. We use the composite score to run a reverse diffusion process $k$ times, generating $k$ candidate images. We then output the image $\hat{I}$ that maximizes the total welfare, as calculated using the agents' submitted bids (which we hope will be close to their private types if we choose a good payment rule). Our allocation rule tries to approximate the VCG allocation rule in the previous section, leveraging our score composition approach to sample over high-welfare parts of $\mathcal{I}$, which is otherwise inaccessible from an optimization standpoint. Our score composition technique is empirically found to generate images that combine the agents' prompts according to the strengths of their bids. So while we cannot optimize welfare exactly over images, our sampling method gives us a good Monte Carlo sampling approach to identify high-welfare images. For some illustrations of our allocation rule see Figures~\ref{fig:grid_images_3_agents} and~\ref{fig:grid_images_2_agents}. 

We also test alternative score composition techniques, namely linearly combining the bidders' scores according to the relative strengths of their bids. However, this leads to worse performance due to diffusion model sensitivity to conflicting gradients. Objects tend to overlay incoherently rather than creating a cohesive image. See Appendix~\ref{app:linear_aggregation} for some samples. This is to say that choosing a good composition method is crucial for our intended application.

\subsection{Payment Rule Alternatives: Myerson Lemma-Based and Clarke Rule-Based}
\label{sec:payment_rule}

We propose two alternatives for the payment rule of the mechanism, one based on Myerson's lemma, and another based on Clarke's payment rule. The former is closer to yielding incentive compatibility of the resulting mechanism but the latter is easier to compute.

\subsubsection{Computing Prices Using Myerson's Lemma.} \label{sec:payments via Myerson's lemma} As discussed in the introduction, the empirical properties of diffusion guidance suggest that the allocation rule of the previous section should be approximately monotone, in that increasing one's bid should result in an expected higher alignment of the sampled image with their caption. This is empirically verified in Section~\ref{sec:alignment-bid}. This suggests a route towards identifying a payment rule that, combined with the allocation rule, would produce a mechanism with strong incentive compatibility properties, based on Myerson's lemma~\cite{myerson1981optimal}. 
Recall that this lemma establishes that, if the allocation rule $x_i(b_i,b_{-i})$ of some mechanism, is monotone non-decreasing in the agent's bid, $b_i$, while holding the bids, $b_{-i}$, of the other bidders fixed, then there exists a unique payment rule (up to a constant shift) that combined with this allocation rule renders truthful bidding a dominant strategy in the resulting mechanism (see also Figure~\ref{fig:myerson's lemma}):
\begin{equation}
    p_i(b_i,b_{-i}) = b_i \cdot x_i(b_i,b_{-i}) - \int_0^{b_i} x_i(z,b_{-i}) \, dz.
    \label{eq:myerson_payment}
\end{equation}
Under this payment rule, an agent with true value $m_i$ who bids truthfully ($b_i = m_i$) obtains utility $u_i(m_i,m_i) = m_i \cdot x_i(m_i,b_{-i}) - p_i(m_i,b_{-i})$, and no alternative bid would yield them a strictly higher utility $u_i(m_i,b_i) = m_i \cdot x_i(b_i,b_{-i}) - p_i(b_i,b_{-i})$. 

To apply this lemma to our setting, we substitute the expected alignment in place of $x_i(b_i,b_{-i})$ i.e.~we take $x_i(b_i,b_{-i}):=\mathbb{E}\left[\alpha(I,c_i)|b_i,c_i,b_{-i},c_{-i}\right]$, where the expectation is taken with respect to running the reverse diffusion process $k$ times and choosing the welfare-maximizing image among the $k$ samples. With this substitution in place, we  notice that, in our setting, we can also write the expected value of a bidder whose private type is $m_i$ and who bids $b_i$ as $m_i \cdot x_i(b_i,b_{-i})$. So Myerson's lemma implies that the price rule~\eqref{eq:myerson_payment} would induce a truthful mechanism, if our allocation rule was indeed monotone non-decreasing. 

We experimentally evaluate the approximate monotonocity of the allocation rule in Section~\ref{sec:alignment-bid} and the approximate truthfulness induced by applying Myerson's lemma for computing prices, in Section~\ref{sec:Myerson-based rule approximate truthfulness experiments}.  As discussed in the Introduction, there are two additional issues we need to consider when computing prices: (1) Our allocation rule is randomized, and in particular we only have sample access to $\alpha(I,c_i)$ for a given set of bids and captions. This is not a big issue. We can apply Myerson's lemma on a sampled curve (whose randomness is with respect to the randomness in running the diffusion process) and for various bids by the bidder (holding the other bidders' bids fixed) and we would get the desired price, in expectation, and therefore maintain the incentive properties ex ante. (2) We do not have an analytical description of the expected curve or a sampled curve to apply Myerson's lemma.  So we will need to do numerical integration to estimate the price, which introduces additional error in incentive compatibility. The errors resulting from the approximate monotonicity and the numerical integration should be small, as illustrated in Figure~\ref{fig:myerson_regret}, and the latter is also controllable with computational effort (increasing the grid points). Everything considered, our experiments suggest that the resulting mechanism has promising incentive compatibility properties. The main disadvantage of the afore-described approach is that it takes time to perform numerical integration, and this  makes the price computation slow. As a heuristic alternative, we propose using prices inspired by the Clarke payment rule, in the next section.

\subsubsection{Computing Prices Using Clarke's Payment Rule.} \label{sec:payments via Clarke's rule} 
To avoid the computational burden of computing prices based on Myerson's lemma, as discussed in the previous section, we propose using a payment rule inspired by Clarke's payment rule in the VCG mechanism. For each bidder $i$, we run our allocation rule without this bidder's bid or caption to sample an image $\hat{I}_{-i}$. We then charge this bidder the difference between the welfare of the other bidders  under this image and under the image $\hat I$ sampled with all bidders included, namely 
\begin{equation}
    p_i = \left( \sum_{j \neq i} v_j(\hat{I}_{-i}) \right) - \left( \sum_{j \neq i} v_j(\hat{I}) \right).
\end{equation}
This is an approximation to the Clarke payment rule~\eqref{eq: Clarke payment rule}, which optimizes (not via Monte Carlo sampling) over all natural images $\mathcal{I}$. Since there is approximation error in the Clarke payment computations, the payment rule will not be guaranteeing incentive compatibility. We perform extensive experimentation to evaluate how approximately truthful the resulting mechanism is in Section~\ref{sec:clarke-based rule approximate truthfulness experiments}, and it shows reasonable performance, albeit it is worse than the performance of the Myerson lemma-based payment rule. The incentive to deviate from truthful bidding is found to decrease with the number $k$ of Monte Carlo samples in the allocation rule. We finally note that, besides the computational efficiency of the Clarke rule-based payments, they offer another advantage, which is protection against the strategic manipulation of captions by the bidders, as they are not tied to the single-dimensional setting we study here, where the captions are public and only the value-per-unit-of-alignment is private information held by the bidders. While we didn't conduct experiments to evaluate how strong the incentives for caption manipulations are, this is a promising direction for future work.



\section{Results}

\subsection{Experimental Setup}

To comprehensively analyze our proposed mechanism, we evaluate properties of the allocation rule and the payment rule, across a diversity of prompts, bids, and the number of bidders, amounted to more than $80,000$ image generations over all. Each setting comprises a base prompt that captures an everyday scenario, ranging from natural backdrops to indoor spaces. These base prompts are paired with bidder prompts in various sectors such as technology, food, fashion, and automobiles, capturing a representative view of modern consumer categories. Table \ref{tab:prompts_2_agents} displays example prompt configurations in two-agent scenarios.

\begin{table}[H]
\captionsetup{skip=2pt}
\caption{Example prompt configurations across 2-Agent scenarios}
\centering
\footnotesize
\begin{tabular}{>{\raggedright\arraybackslash}p{6cm} 
                >{\raggedright\arraybackslash}p{3cm} 
                >{\raggedright\arraybackslash}p{3cm}}
\toprule
\textbf{Base Prompt} & \textbf{Agent 1 Prompt} & \textbf{Agent 2 Prompt} \\
\midrule
Two friends chatting over coffee at a cafe & Starbucks mug & Apple MacBook Air \\ \hline
A hiker resting on a mountain trail & North Face backpack & Gatorade bottle \\ \hline
A woman with a tray of food at a cafeteria & McDonald's fries & Pepsi can \\ \hline
A person walking down a city street & Louis Vuitton handbag & H\&M storefront \\ \hline
Friends gathered for a movie night at home & Doritos chips & Coca-Cola cans \\ \hline
A car driving along a scenic coastal highway & Tesla car & ExxonMobil sign \\
\bottomrule
\end{tabular}
\label{tab:prompts_2_agents}
\end{table}

For example, the following prompts correspond to the generations from our allocation rule, shown in Figure \ref{fig:grid_images_2_agents}, under different combinations of bids for two agents:

\begin{figure}[tb]
    \centering
    \begin{subfigure}[b]{0.193\textwidth}
        \centering
        \includegraphics[width=\textwidth]{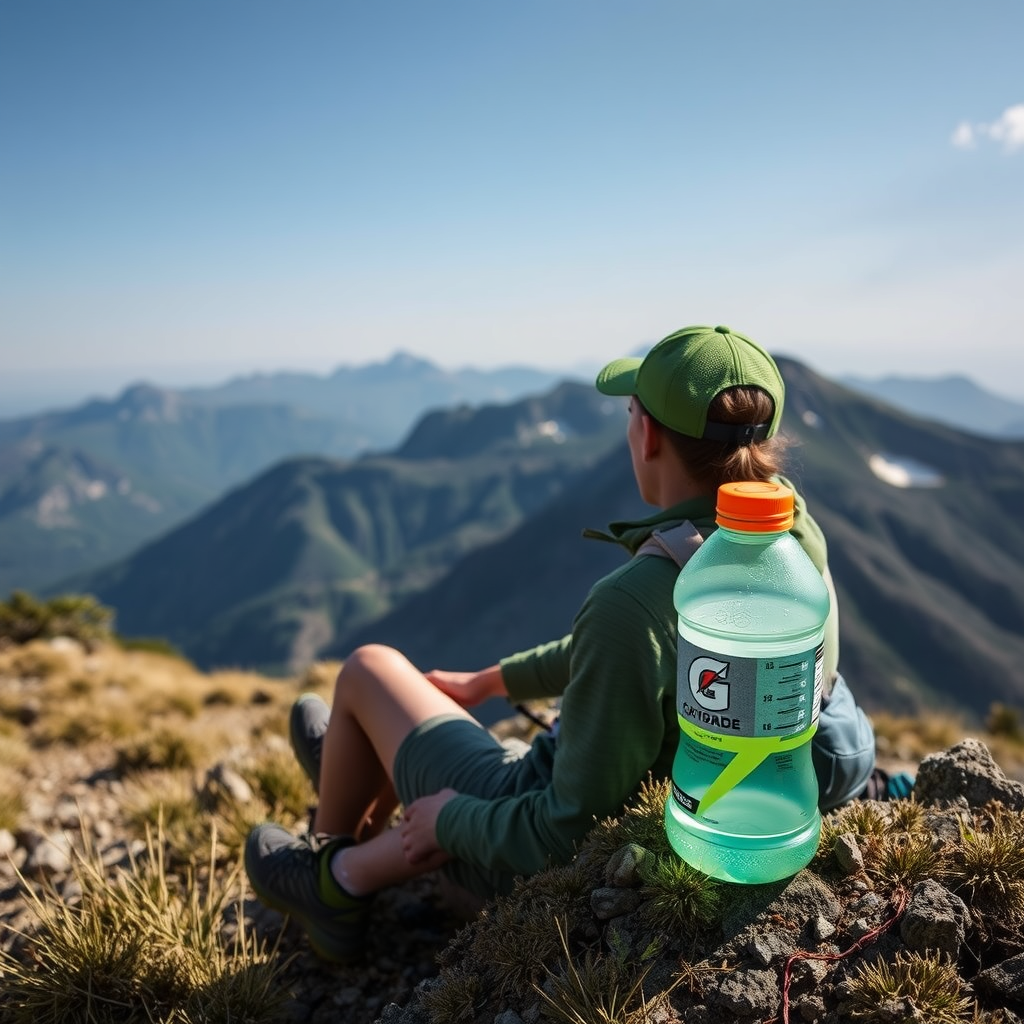}
        \caption{$(0, 1)$}
        \label{fig:sub1}
    \end{subfigure}
    \begin{subfigure}[b]{0.193\textwidth}
        \centering
        \includegraphics[width=\textwidth]{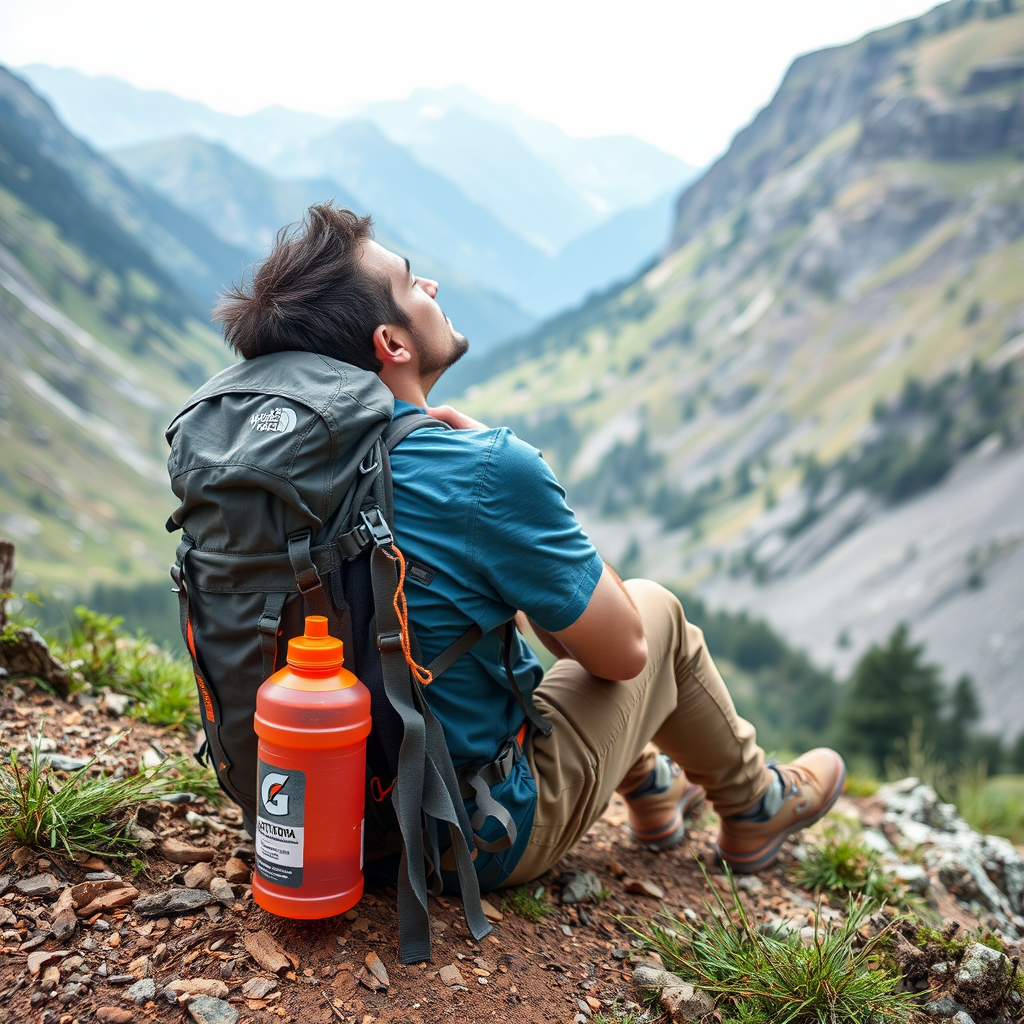}
        \caption{$(0.3, 0.7)$}
        \label{fig:sub2}
    \end{subfigure}
    \begin{subfigure}[b]{0.193\textwidth}
        \centering
        \includegraphics[width=\textwidth]{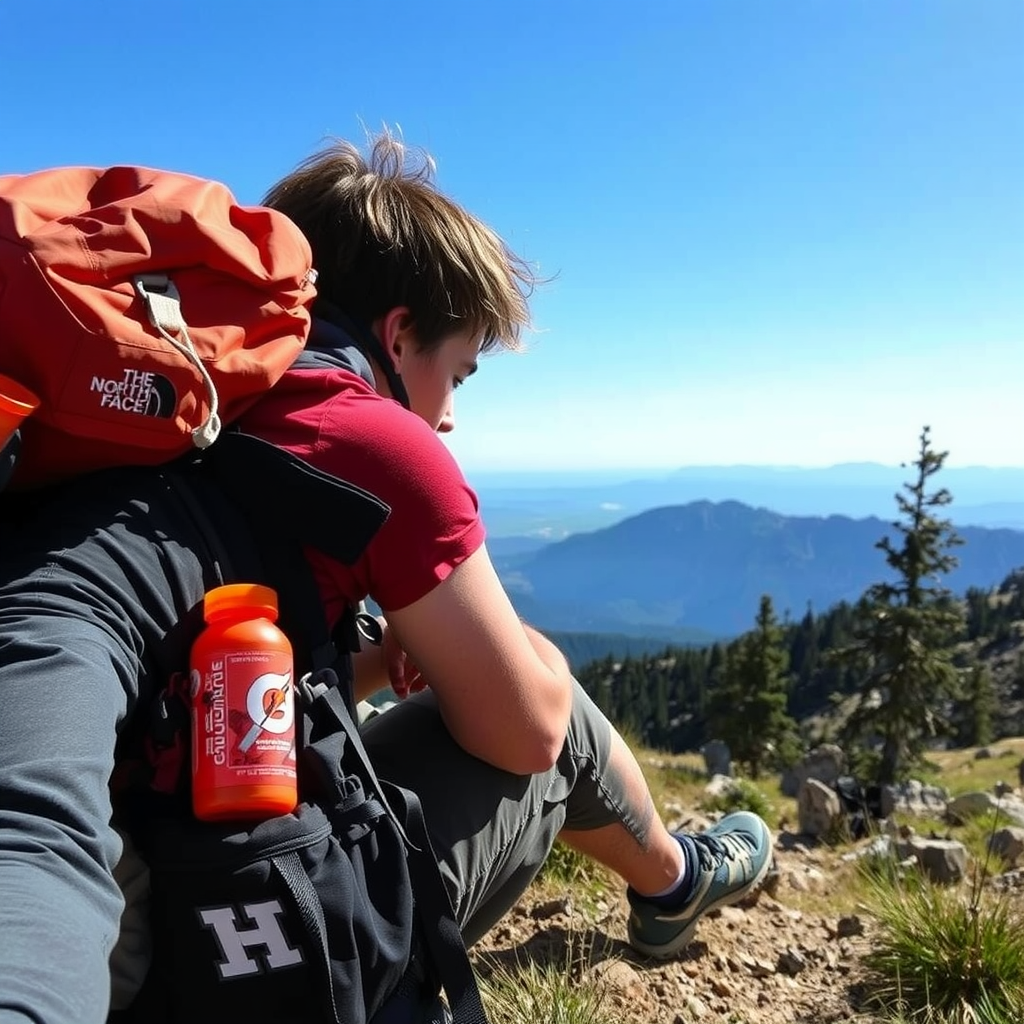}
        \caption{$(0.5, 0.5)$}
        \label{fig:sub3}
    \end{subfigure}
    \begin{subfigure}[b]{0.193\textwidth}
            \centering
            \includegraphics[width=\textwidth]{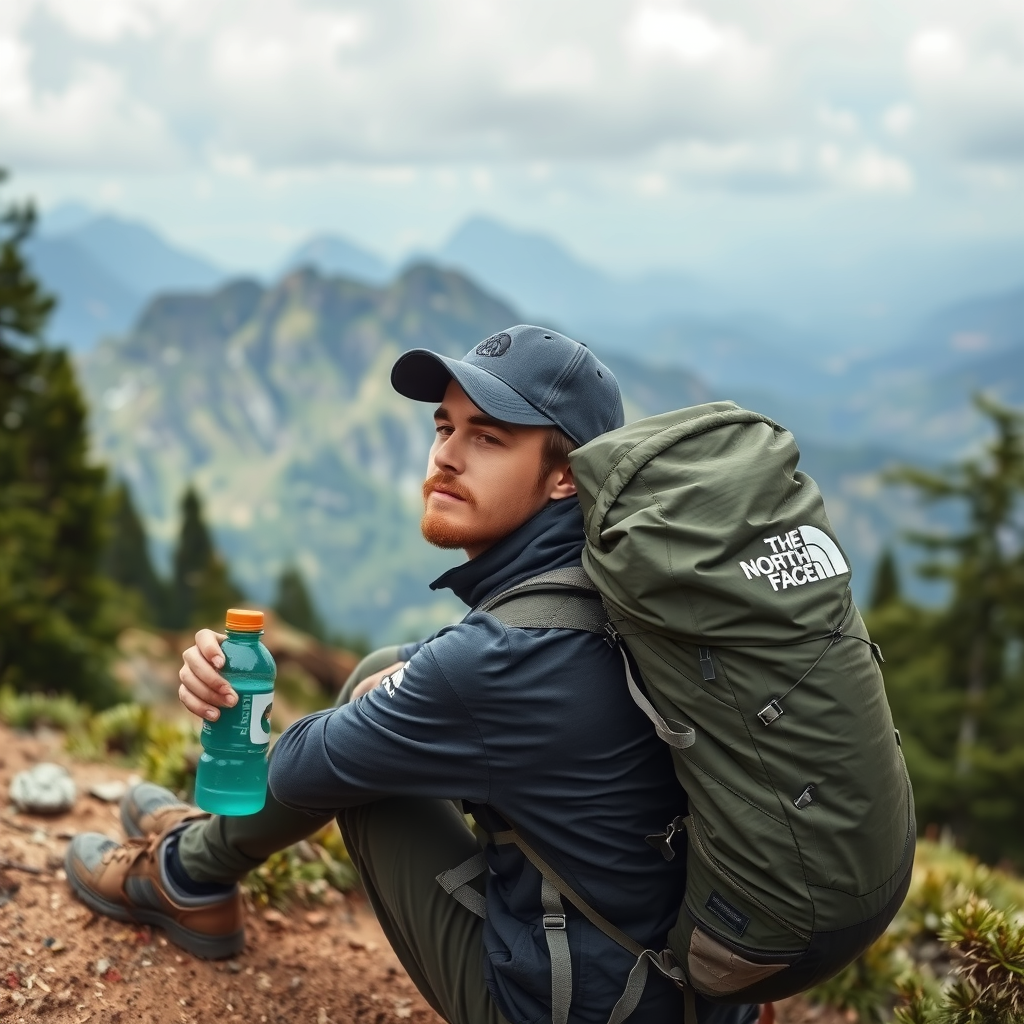}
            \caption{$(0.7, 0.3)$}
            \label{fig:sub4}
        \end{subfigure}
    \begin{subfigure}[b]{0.193\textwidth}
        \centering
        \includegraphics[width=\textwidth]{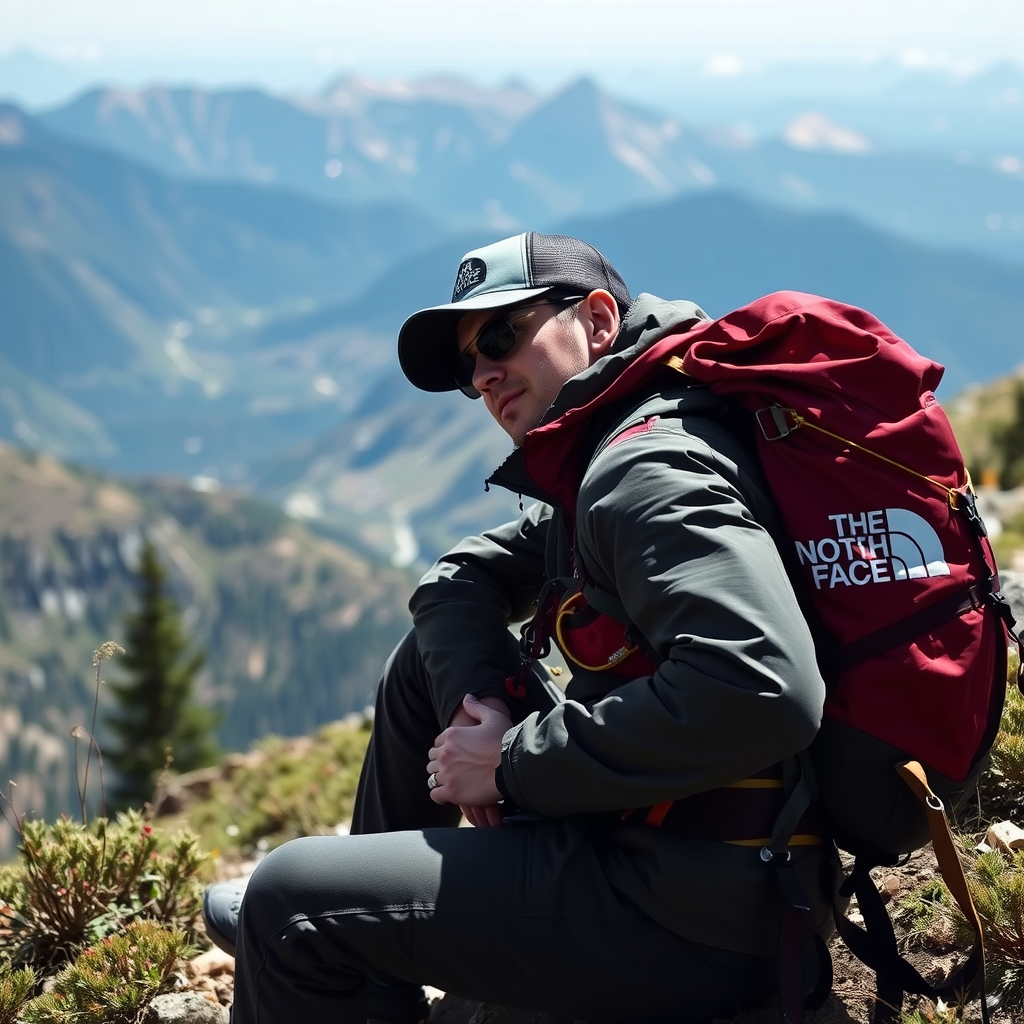}
        \caption{$(1, 0)$}
        \label{fig:sub5}
    
    \end{subfigure}
    
    \caption{Allocation rule outputs for 2 agents, North Face and Gatorade, with bids $(b_1, b_2)$.}
    \label{fig:grid_images_2_agents}
\end{figure}

\begin{enumerate}[label=\(\bullet\), leftmargin=6mm, nosep]
\itemsep0.10cm
    \item \textbf{Base:} ``A hiker resting on a mountain trail"
    \item \textbf{Base + Agent 1:} ``A hiker resting on a mountain trail with North Face backpack"
    \item \textbf{Base + Agent 2:} ``A hiker resting on a mountain trail with Gatorade bottle"
    \item \textbf{Base + Joint:} ``A hiker resting on a mountain trail with North Face backpack and Gatorade bottle"
\end{enumerate}

In the next sections, we describe our findings from experimenting with our proposed allocation rule and mechanism across various settings, involving different prompts and bid combinations. We test the robustness of our findings by considering settings involving $2$, $3$, $5$, $10$, and $20$ bidders.  For example,  Figure~\ref{fig:grid_images_3_agents} displays example outputs of our allocation rule for $3$ agents.

Appendix~\ref{app:prompts} details the full prompt generation setup and Appendix~\ref{app:bid-dist} shows the distribution of bids that we sweep over. Our setup allows us to systematically vary agents' bids under various prompts, allowing a range of score compositions. As per the definition of our allocation rule, for each combination of bids and prompts, we conduct Monte Carlo sampling, generating $k$ images and output the welfare-maximizing one. All images are generated using the \texttt{FLUX.1-schnell} diffusion model with 5 inference steps and a guidance scale of 10. To compute welfare, we use CLIP \citep{radford2021clip}. In particular, we compute the alignment $\alpha_i = \operatorname{CLIP}(I, c_i)$ as the cosine similarity between the CLIP image embedding and text embedding, which lie in a shared embedding space.


\subsection{Bid Monotonicity: Prompt Alignment Increases with Bid}
\label{sec:alignment-bid}


\begin{figure}[t]
    \centering
    \begin{subfigure}[b]{0.48\textwidth}
        \centering
        \includegraphics[width=\textwidth]{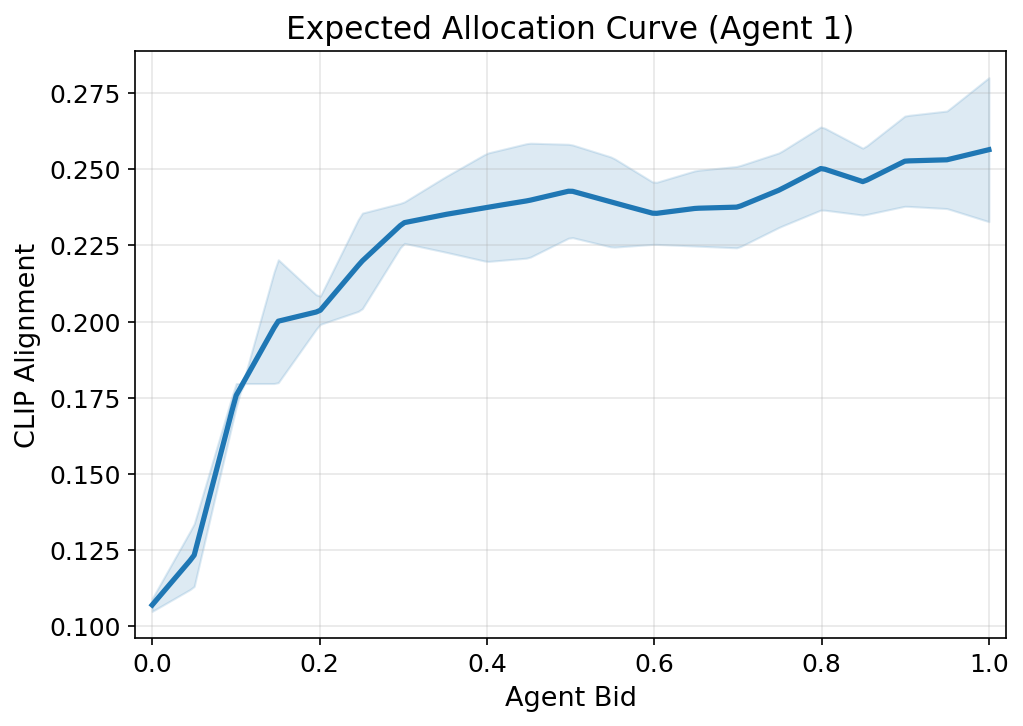}
        \caption{Agent~1's alignment with the output image as her bid is varied.}
        \label{fig:bid_monotonicity_agent1}
    \end{subfigure}
    \hfill
    \begin{subfigure}[b]{0.48\textwidth}
        \centering
        \includegraphics[width=\textwidth]{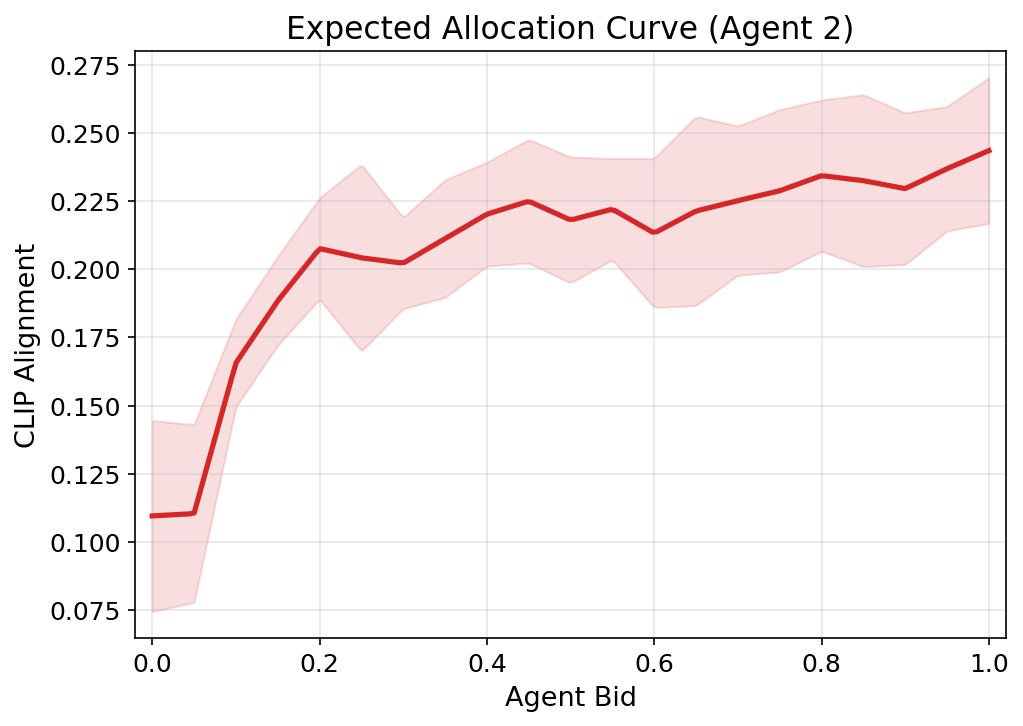}
        \caption{Agent~2's alignment with the output image as her bid is varied.}
        \label{fig:bid_monotonicity_agent2}
    \end{subfigure}
    \caption{Alignment of the image output by the allocation rule with a particular bidder's caption as they vary their bid from $0$ to $1$ in increments of $0.05$, while keeping the opposing agent's bid fixed at $0.3$. For each bid combination and prompt configuration, $k = 5$ images are generated per trial across $5$ trials; the welfare-maximizing image is selected within each trial, and CLIP alignment scores are averaged across the $5$ trials and the two prompts. Shaded regions indicate standard deviation. }
    \label{fig:bid_monotonicity}
\end{figure}

An important property that we would like our allocation rule to have is \emph{bid monotonicity}: agents who bid higher should receive higher allocation, meaning that the image selected by the allocation rule should have a higher expected alignment with their prompt as their bid increases. To study this, we consider two representative prompt combinations in the two-agent setting.\footnote{Specifically, one setting uses ``A woman with a tray of food at a cafeteria,'' as base prompt, with ``McDonald's fries'' and ``Pepsi can,'' as prompts for Agents 1 and 2. The other setting uses ``A person walking down a city street,'' as base prompt, and ``Louis Vuitton handbag'' and ``H\&M storefront,'' as prompts for Agents 1 and 2.} For each of these prompt configurations, we fix one of the two agent's bid at $b = 0.3$ and vary the other agent's bid over $\{0.00, 0.05, 0.10, \ldots, 1.00\}$. For each of the 21 bid combinations, we run our allocation rule using $k = 5$, i.e.~we generate $5$ images using our composite score and select the welfare-maximizing one. To obtain a better estimate of the expected alignment, we repeat $5$ times, and compute the average CLIP alignment across the $5$ repetitions. 

Figure~\ref{fig:bid_monotonicity} shows the curves plotting the average allocation to each agent as they increase their bid from $0$ to $1$, holding the other bidder's bid fixed at $0.3$. We show the average of Agent 1's and Agent 2's allocation, averaging across repetitions and the two prompt configurations. The shaded region represents the standard deviation across samples. Figure~\ref{fig:myerson_regret} (top plots) shows  Agent 1's allocation split between each of the prompt configurations. These curves are approximately monotone non-decreasing, consistent with our allocation rule satisfying approximate bid monotonicity.
This is also visualized in Figures~\ref{fig:grid_images_3_agents} and~\ref{fig:grid_images_2_agents}.

\subsection{Efficiency: Improved Welfare Compared to Single-Winner Baseline}
\label{sec:efficiency}

\begin{figure}[tb]
  \centering
  \begin{subfigure}[b]{0.43\textwidth}
      \centering
      \includegraphics[width=\textwidth]{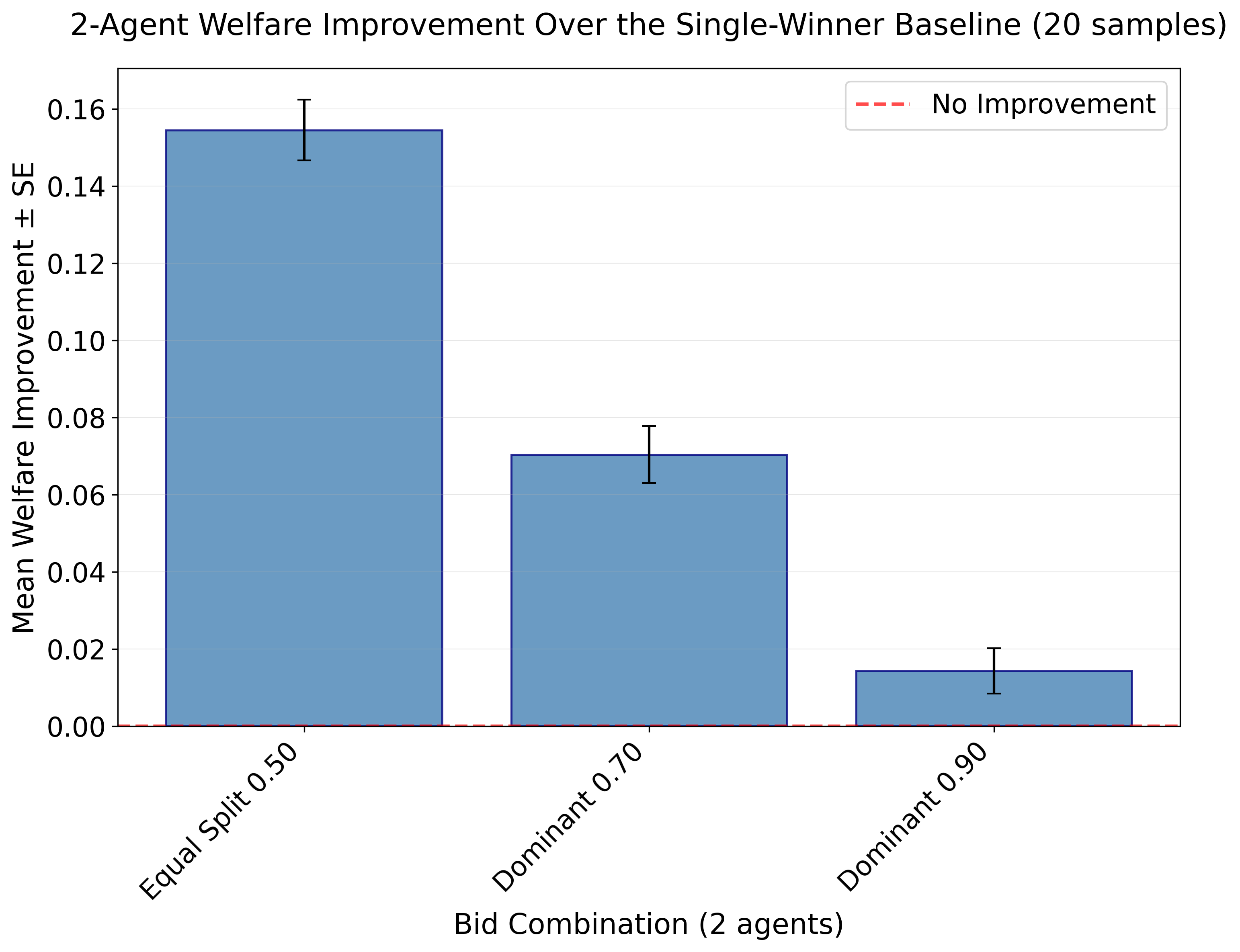}
      \caption{Average welfare improvement of up to 15.4\% over single-winner baseline for 2 agents.}
      \label{fig:welfare_over_bids_2_agent}
  \end{subfigure}
  \hspace{1cm}
  \begin{subfigure}[b]{0.43\textwidth}
    \centering
    \includegraphics[width=\textwidth]{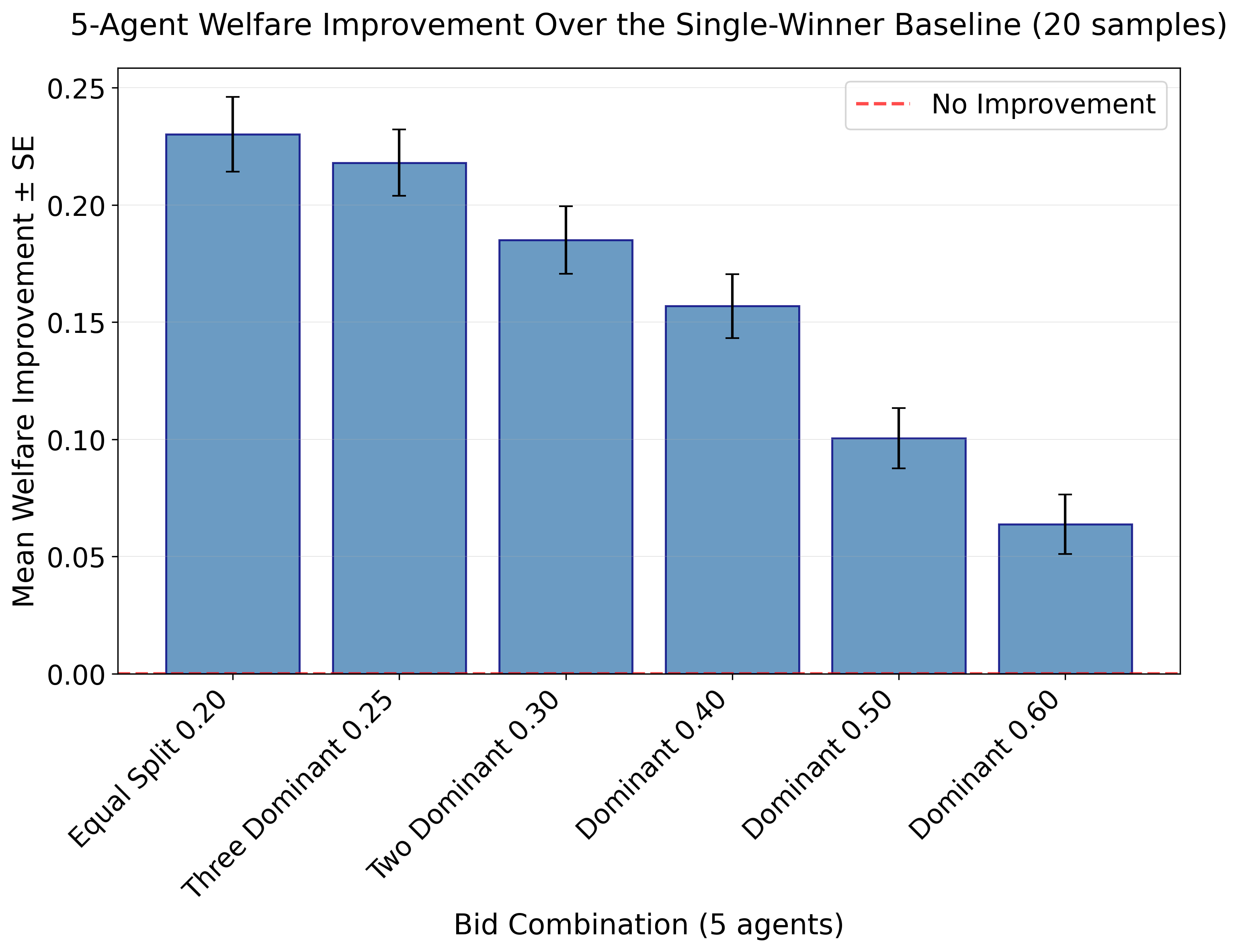}
    \caption{Average welfare improvement of up to 23.0\% over single-winner baseline for 5 agents.}
    \label{fig:welfare_over_bids_5_agent}
  \end{subfigure}
  \caption{Percent improvement of the welfare from our allocation rule over the single-winner baseline, averaged across all prompts in the two-agent and the five-agent settings respectively. We use $k=20$ Monte Carlo samples in the allocation rule, for each prompt and bid combination.}
  \label{fig:welfare_across_bids}
\end{figure}

Unlike single-winner allocations, our allocation rule enables multiple agents to be represented simultaneously in the generated image. It is important to evaluate whether this compositional capability of our allocation rule leads to a higher total welfare.

\paragraph{Single-Winner Baseline.}
We compare our allocation rule against a single-winner baseline that generates images considering a single agent's prompt at a time. In this baseline, for a number of bidders $n$ and parameter $k$ in the allocation rule, for each combination of a base prompt and bidder prompts, $n$ images are generated from our allocation rule, using the base prompt and a single agent's prompt; in particular, for each agent, $k$ images are generated using this agent's prompt appended to the base prompt and the image that maximizes welfare for this agent is chosen. After $n$ images are generated, one corresponding to each bidder's prompt, the final output  is the image that optimizes the overall welfare. This baseline serves as a direct comparison that lacks compositionality of agent preferences.

\paragraph{Results.}
Our allocation rule consistently outperforms the single-winner allocation rule described above, across our experiments, varying the agent prompts and the number of agents.

For the $2$-agent setting, Figure~\ref{fig:welfare_over_bids_2_agent} shows that our allocation rule improves the average welfare (averaged over all $200$ prompts we consider in the two-agent setting) by up to $15.4$\% over the baseline, which occurs when both agents' normalized bids are equal to $0.5$. In the $5$-agent setting, Figure~\ref{fig:welfare_over_bids_5_agent} shows that our allocation rule improves the average welfare (averaged over all $85$ prompts we consider in the five-agent setting) by up to $23.0$\% over the baseline, which occurs when all 5 agents' normalized bids are equal to $0.2$. Welfare improvements for up to $20$ agents are included in Appendix \ref{app:welfare_all_agents}.

\begin{wrapfigure}[14]{r}{0.5\textwidth}
\vspace{-0.5cm}
\centering
    \includegraphics[width=0.4\textwidth]{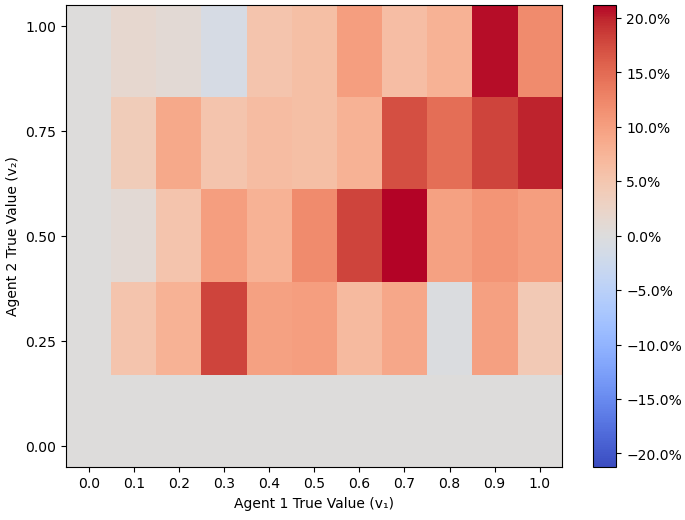}
    \caption{Welfare improvement is most significant when agents bid similar values.}
    \label{fig:welfare_heatmap}
\end{wrapfigure}

Individual prompts exhibit even larger improvements; e.g., we see up to $21$\% welfare improvement in the two-agent setting and up to $36.0$\% welfare improvement in the $5$-agent setting, for specific prompt and bid combinations. In Figure \ref{fig:welfare_heatmap} we show results when the base prompt is ``A car driving along a scenic coastal highway,'' and the bidders' prompts are ``Tesla car'' and ``ExxonMobil sign.'' 
In this plot we use $k=50$ Monte Carlo samples in the allocation rule, and vary Agent 1's bid from 0.0 to 1.0 in increments of 0.1 and Agent 2's bid from 0.0 to 1.0 in increments of 0.25. The welfare improvement over the baseline is the highest when the agents bid similar values. This occurs because the bidders' prompts are compatible and balanced bids activate the cooperative potential in our score composition. This allows the allocation rule to sample high-welfare outcomes that are  inaccessible to the single-winner allocation rule, which is limited to using one agent's prompt at a time.

\begin{wrapfigure}[15]{r}{0.5\textwidth}
\centering
    \includegraphics[width=0.45\textwidth]{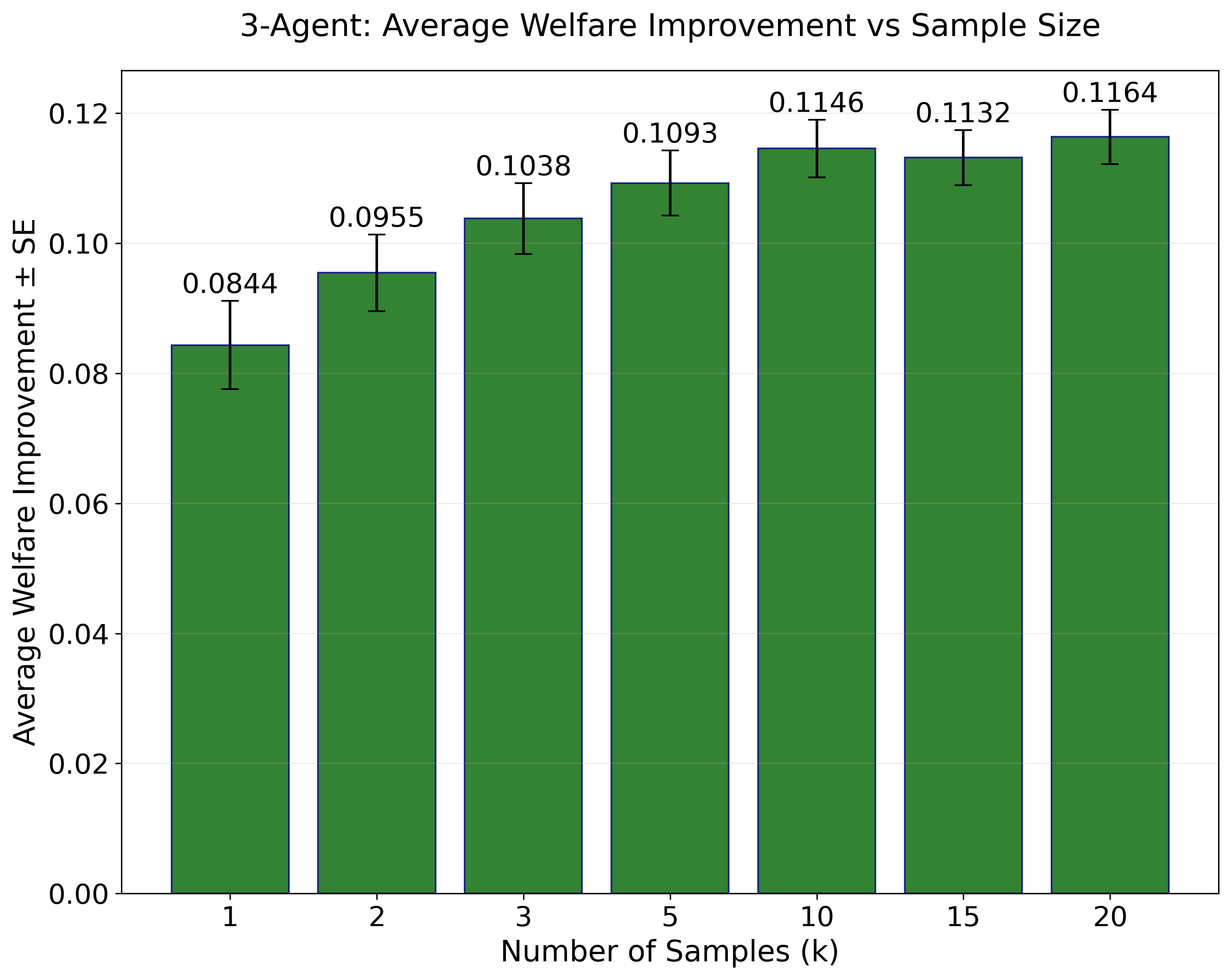}
    \caption{$k=5$ Monte Carlo samples in the allocation rule attain more than 90\% of the welfare improvement over the single-winner baseline attained with $k=20$.}
    \label{fig:welfare_across_k_3_agents}
\end{wrapfigure}

Finally, we conduct experiments varying the number $k$ of Monte Carlo samples in the allocation rule, from $k=1$ to $k=20$ and for varying numbers of agents. For up to 10 agents, we see a rapid convergence of the welfare improvement over the single-winner baseline, with $k = 5$ samples achieving at least 90\% of the welfare improvement attained with $k=20$ samples. Figure \ref{fig:welfare_across_k_3_agents} shows the average, across our 3-agent settings, welfare improvement as a function of $k$. Welfare improvement plateaus beyond $k = 5$, indicating that the benefits from Monte Carlo sampling saturate quickly. For 20 agents,  $k = 15$ achieves a similar effect. 
Our detailed results 
are included in Appendix \ref{app:mc_all_agents}.
For real-time advertising scenarios where latency matters, these plots suggest that the benefits from our allocation rule are realized for moderate values of $k$.

\subsection{Incentive Compatibility}
\label{sec:truthfulness}
We evaluate the incentive properties of our mechanism under the two payment rules we proposed.
\subsubsection{Near Truth-Inducing Payments from Myerson's Lemma} \label{sec:Myerson-based rule approximate truthfulness experiments}

The approximate bid monotonicity of the allocation rule suggests a path towards a payment rule that induces an approximately truthful mechanism, based on Myerson's lemma~\cite{myerson1981optimal}, as developed in Section~\ref{sec:payments via Myerson's lemma}.

In this section, we experimentally evaluate the impact to incentive compatibility from the approximate monotonicity of the allocation rule and the numerical integration involved in computing Myerson payments. 
To estimate the ex ante regret of the agents for not reporting truthfully, we approximate the (ex ante) allocation of the auction, $x_i(b_i,b_{-i}):=\mathbb{E}\left[\alpha(I,c_i)|b_i,c_i,b_{-i},c_{-i}\right]$, by replacing $\mathbb{E}$ with the empirical average over $5$ trials. Using the approximate allocation rule $\hat x_i$, we compute approximate Myerson payments $\hat p_i$, by applying the trapezoidal rule on  Eq.~\eqref{eq:myerson_payment} (with $x_i$ replaced by $\hat x_i$). From approximate allocations and prices, we compute approximate utilities $\hat u_i(m_i, b_i)=m_i \cdot \hat x_i(b_i,b_{-i}) - \hat p_i(b_i,b_{-i})$ for bidding $b_i$ when the true type is $m_i$, and finally compute Agent 1's \emph{relative regret} from truthful bidding:
\begin{equation}
    \text{Regret}(m_i) = \frac{\max_{b_i} \, \hat u_i(m_i, b_i) - \hat u_i(m_i, m_i)}{|\hat u_i(m_i, m_i)|},
\end{equation}
maximizing over all bids on a grid.

\begin{figure}[t]
    \centering
    \begin{subfigure}[b]{0.48\textwidth}
        \centering
        \includegraphics[width=\textwidth]{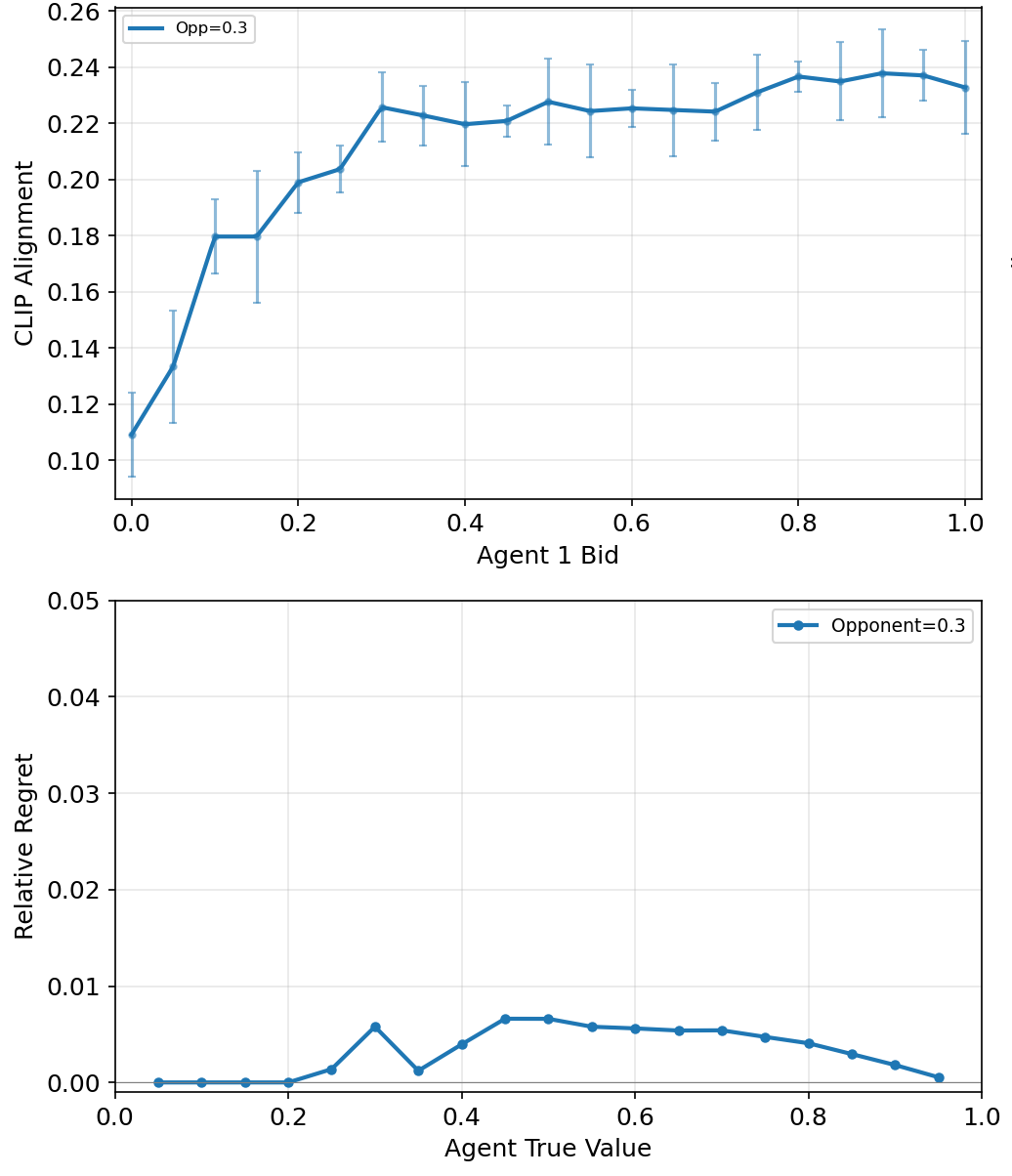}
        \caption{Base: A woman with a tray of food at a cafeteria, Agent~1: McDonald's fries, Agent~2: Pepsi can.}
        \label{fig:myerson_p1}
    \end{subfigure}
    \hfill
    \begin{subfigure}[b]{0.48\textwidth}
        \centering
        \includegraphics[width=\textwidth]{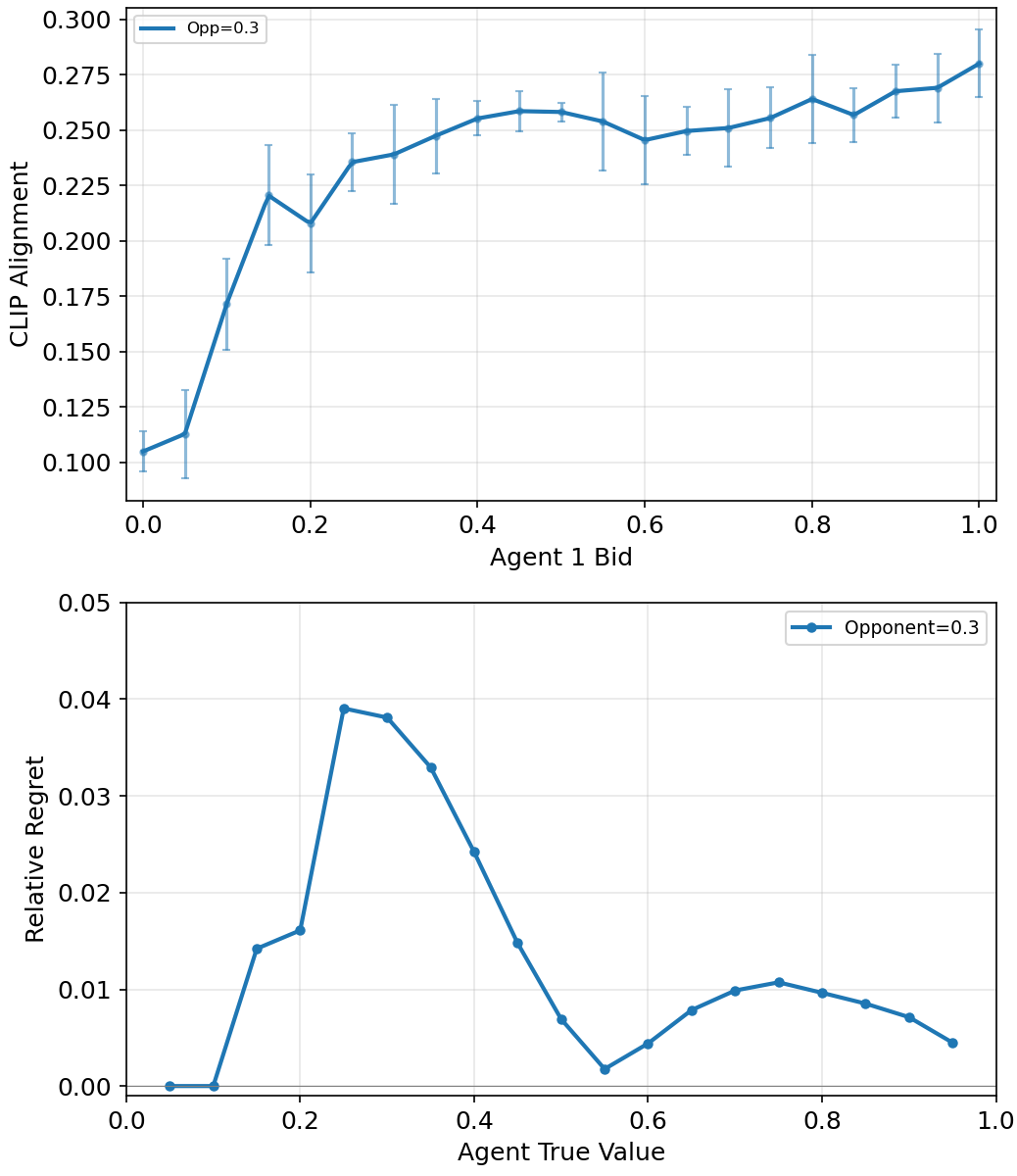}
        \caption{Base: A person walking down a city street, Agent~1: Louis Vuitton handbag, Agent~2: H\&M storefront.}
        \label{fig:myerson_p2}
    \end{subfigure}
    \caption{Allocation rule (top) and relative regret from truthful bidding (bottom) under the Myerson payment rule, with Agent~2's bid fixed at $b_2 = 0.3$ and Agent~1's bid varied over $[0, 1]$ in increments of $0.05$. The allocation rule uses $k = 5$ diffusion samples. To estimate the expected alignment induced by our allocation rule we sample from the allocation rule $5$ times and average the alignment scores from these $5$ repetitions. Relative regret stays below $4\%$ across all true values, confirming near incentive compatibility.}
    \label{fig:myerson_regret}
\end{figure}

Figure~\ref{fig:myerson_regret} shows the results for two representative prompts in a two-agent setting. As seen earlier, the allocation curves are approximately monotone non-decreasing, confirming the precondition of Myerson's lemma. The relative regret remains below 4\% across all tested true values $m_i \in \{0.1, 0.2, \ldots, 0.9\}$, with most values much lower, indicating that the mechanism with Myerson payments has very good incentive properties. 

\subsubsection{Approximate Truthfulness from Clarke-Based Payments} \label{sec:clarke-based rule approximate truthfulness experiments} 
As discussed in Section~\ref{sec:payment_rule}, a main disadvantage of payments computed via Myerson's lemma is that it takes time to perform numerical integration of the allocation curve. This motivated us to propose an alternative payment rule based on Clarke's payment rule in the VCG mechanism, in Section~\ref{sec:payments via Clarke's rule}. In this section, we estimate the regret of the agents for not reporting truthfully. Since Clarke payments are faster to compute, we also study the ex post regret of the agents.

\begin{figure}[t]
    \centering
    \begin{subfigure}[b]{0.48\textwidth}
        \centering
        \includegraphics[width=\textwidth]{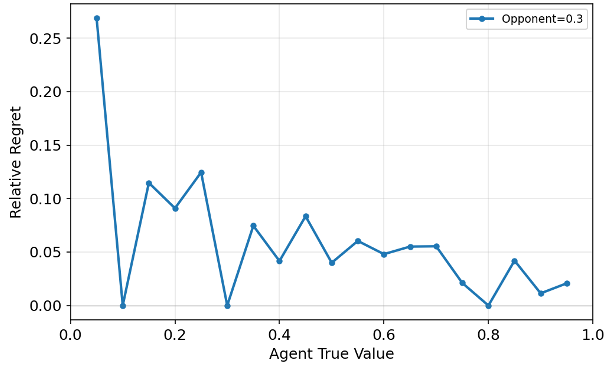}
        \caption{Base: A woman with a tray of food at a cafeteria, Agent~1: McDonald's fries, Agent~2: Pepsi can.}
        \label{fig:clarke_p1}
    \end{subfigure}
    \hfill
    \begin{subfigure}[b]{0.48\textwidth}
        \centering
        \includegraphics[width=\textwidth]{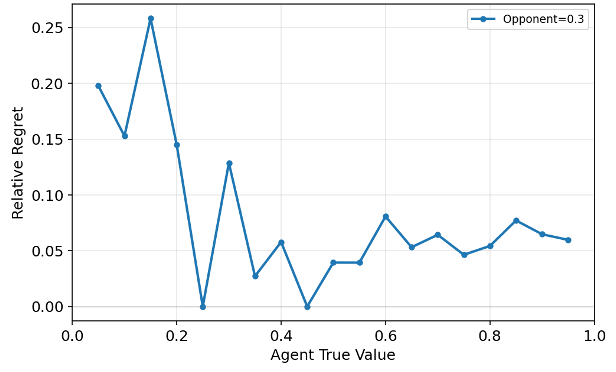}
        \caption{Base: A person walking down a city street, Agent~1: Louis Vuitton handbag, Agent~2: H\&M storefront.}
        \label{fig:clarke_p2}
    \end{subfigure}
    \caption{Relative regret from truthful bidding under the Clarke payment rule for the exact same settings as those in Figure~\ref{fig:myerson_regret}. The allocation rule uses $k = 5$ diffusion samples. To estimate the (ex ante) Clarke payments and utilities, we run the allocation rule $5$ times and average the results. The relative regret is worse than that under Myerson lemma-based payments, by about an order of magnitude.}
    \label{fig:clarke_regret}
\end{figure}

\paragraph{Ex Ante Regret} In Figure~\ref{fig:clarke_regret}, we plot our estimates of the bidders' (ex ante) regrets for not bidding truthfully when Clarke payments are charged, in the same settings as those of Figure~\ref{fig:myerson_regret}. To estimate the (ex ante) Clarke payments and utilities, we run the allocation rule $5$ times and average the results. We use $k=5$ for the allocation rule. The relative regrets are found to be controlled, for the most part below $10$\%, but they are about an order of magnitude worse than those guaranteed by Myerson lemma-based payments. 

\begin{figure}[htb]
  \centering
  \begin{subfigure}[b]{0.48\textwidth}
    \centering
    \includegraphics[width=\textwidth]{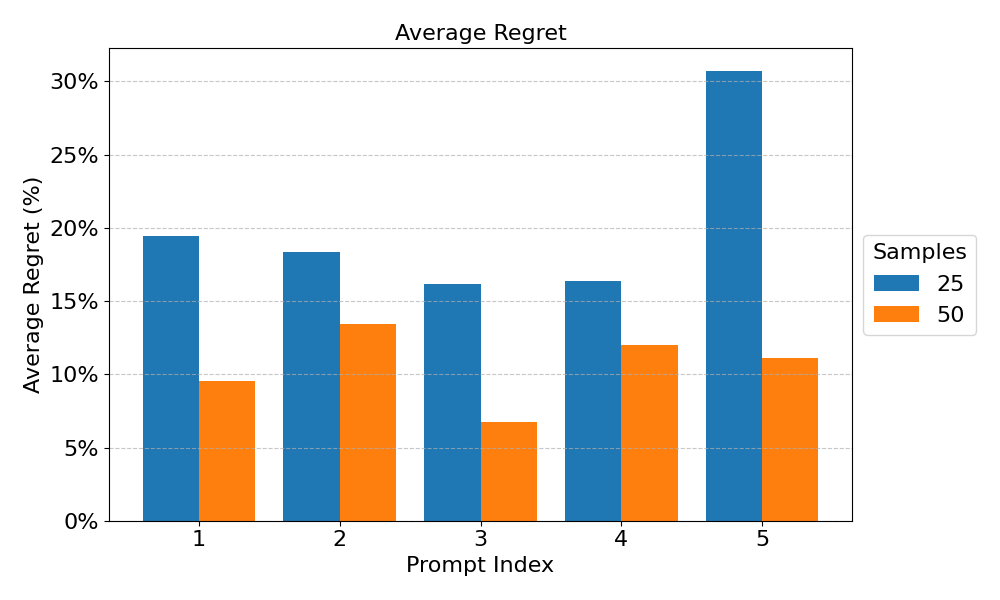}
    \caption{Relative ex post regret.}
    \label{fig:regret_avg}
  \end{subfigure}
  \hfill
  \begin{subfigure}[b]{0.48\textwidth}
    \centering
    \includegraphics[width=\textwidth]{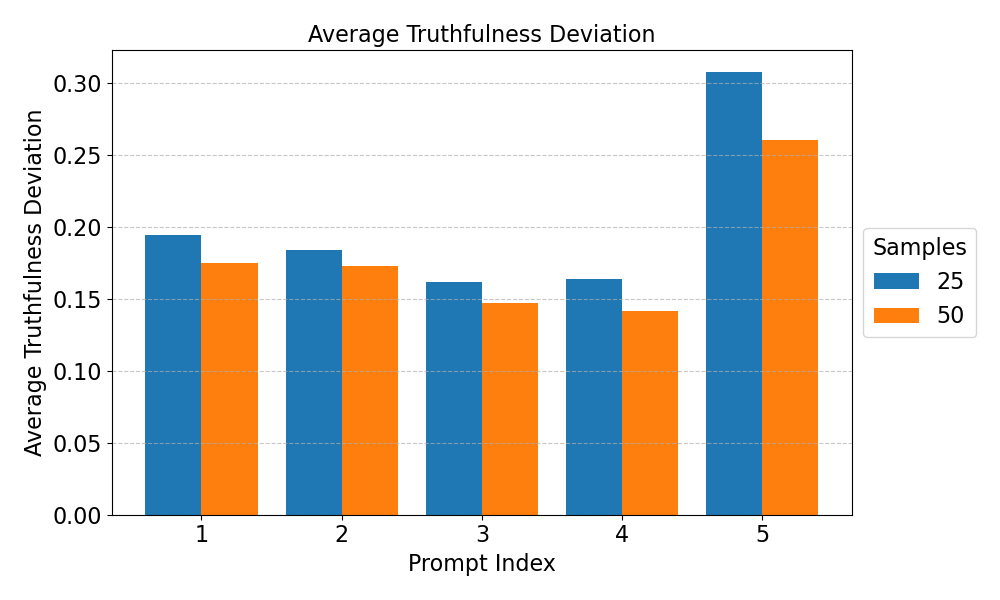}
    \caption{Absolute ex post regret. }
    \label{fig:truth_deviation}
  \end{subfigure}
  \caption{Relative and absolute ex post regret of Agent 1, for $k=25$ and $k=50$ Monte Carlo samples in the allocation rule, for 5 different prompt configurations in the two-agent setting, and averaged over possible values of the agents in a grid over $[0,1] \times [0,1]$.}
  \label{fig:truthfulness_combined}
\end{figure}

\paragraph{Ex Post Regret.}  We also study the ex post regret experienced by the bidders, i.e.~the regret they experience when the allocation rule is run once to select an image, and is run another one time per bidder to estimate their Clarke payments. 

\begin{wrapfigure}[18]{r}{0.5\textwidth}
    \centering
    \vspace{-0.9cm}
    \includegraphics[width=0.45\textwidth]{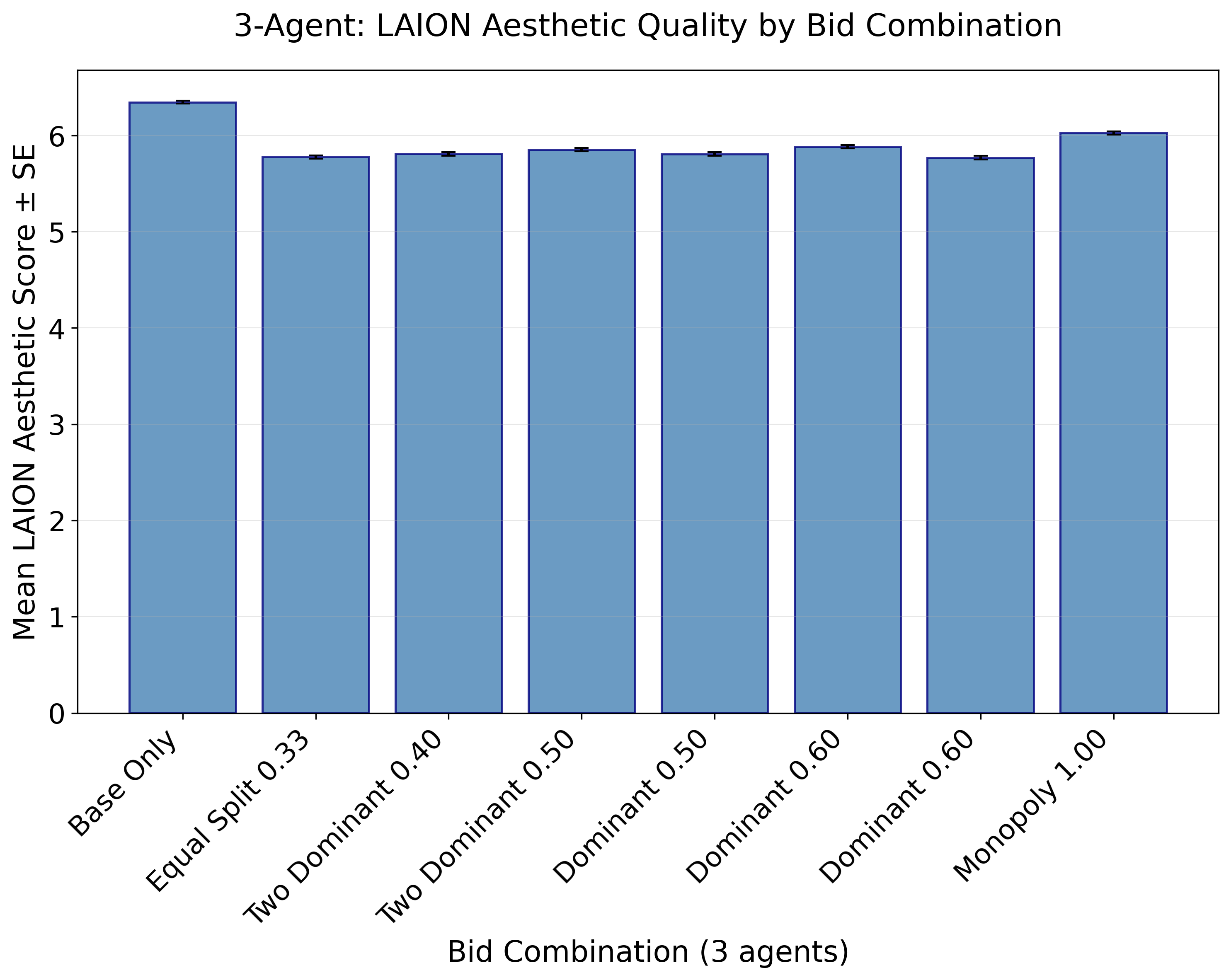}
    \vspace{-0.55cm}
    \caption{LAION image quality scores in the 3-agent setting. Scores are shown for different bid configurations for the agents. Scores are averaged across all the prompt configurations we consider in Appendix~\ref{app:prompts}.}
    \label{fig:laion_quality_3agents_main}
\end{wrapfigure} 
Across five prompt configurations in the two-agent setting, we sweep Agent~1’s true value from 0.0 to 1.0 in steps of 0.1 and Agent~2’s true value from 0.0 to 1.0 in steps of 0.25. For every combination of true values $(m_1, m_2)$, we fix agent 2's bid to be truthful and allow agent 1 to bid a range of values. Figure~\ref{fig:regret_avg} shows the relative regret experienced by Agent~1 across the five prompt configurations, averaged across realized bidder types. We report results for $k\in\{25,50\}$ Monte Carlo samples in the allocation rule. For $k=50$, the average relative regret remains below 15\% across all prompts, reaching as low as 7\% for individual prompts.

Figure~\ref{fig:truth_deviation} reports the averaged, across realized bidder types, absolute regret, i.e.~the difference of Agent~1’s utility when they bid their utility-maximizing bid vs their true value, across five prompts. For $k=50$, the average truthfulness deviation decreases to as low as 0.14. 



\subsection{Image Quality Preservation}



A possible concern from the composition of multiple agent preferences in our allocation rule is that this may result in image quality degradation. To quantitatively evaluate this, we use  LAION Aesthetic Score, a metric trained on human aesthetic~preferences~from~web-scale data~\citep{schuhmann2022laion}.

LAION scores, averaged across all prompt configurations we consider in this paper (see Appendix~\ref{app:prompts}),  were consistently above 5.75, which  indicates high image quality, for all settings we considered in terms of number of agents and bid configurations. For $n=3$ agents, our allocation rule achieves scores ranging from 5.77 to 6.02; see Figure~\ref{fig:laion_quality_3agents_main}.
Full LAION image quality scores for up to 20 agents are shown in Appendix \ref{app:laion}. As an alternative, we also evaluate image quality via CLIP alignment, shown in Appendix \ref{app:clip-alignment}.


\subsection{Competitive Bidding}

We conducted additional experiments specifically testing our mechanism with semantically similar, directly competing products. We evaluate 20 scenarios where two agents bid for competing brands in the same product category (e.g., Dunkin vs. Starbucks, see Table ~\ref{tab:prompts_2_agents_competitive_example}). These scenarios are interesting to test in terms of whether there still exist welfare improvement opportunities. They also represent challenging cases as embedding space distinctions are inherently weaker than our original diverse prompt set. Comprehensive prompts are included in Appendix \ref{app:competitive_bidding}.

\begin{table}[H]
\captionsetup{skip=2pt}
\caption{Example prompt configurations in 2-Agent competitive scenarios}
\label{tab:prompts_2_agents_competitive_example}
\centering
\footnotesize
\begin{tabular}{>{\raggedright\arraybackslash}p{6cm} 
                >{\raggedright\arraybackslash}p{3cm} 
                >{\raggedright\arraybackslash}p{3cm}}
\toprule
\textbf{Base Prompt} & \textbf{Agent 1 Prompt} & \textbf{Agent 2 Prompt} \\
\midrule
Someone drinking from a coffee cup in an office meeting & Starbucks coffee cup & Dunkin coffee cup \\ \hline
A person drinking from a beverage can while working at a desk & Coca-Cola can & Pepsi can \\ \hline
Someone wearing athletic shoes while jogging in a park & Nike running shoes & Adidas running shoes \\
\bottomrule
\end{tabular}
\end{table}

\begin{figure}[H]
    \centering
    \begin{subfigure}[b]{0.475\linewidth}
        \centering
        \includegraphics[width=\linewidth]{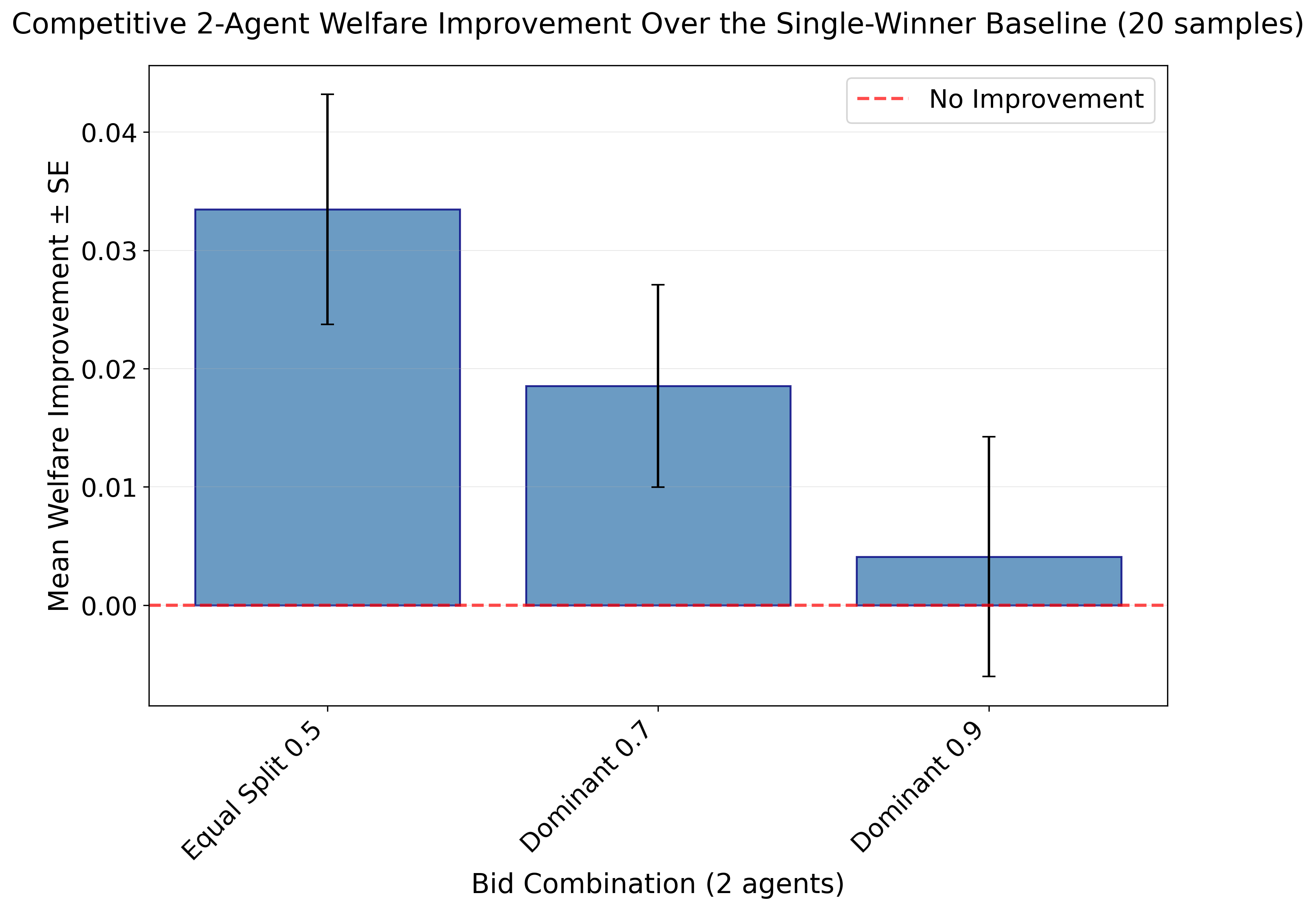}
        \caption{Welfare improvement over baseline}
        \label{fig:welfare_2agent_competitive}
    \end{subfigure}
    \hfill
    \begin{subfigure}[b]{0.46\linewidth}
        \centering
        \includegraphics[width=\linewidth]{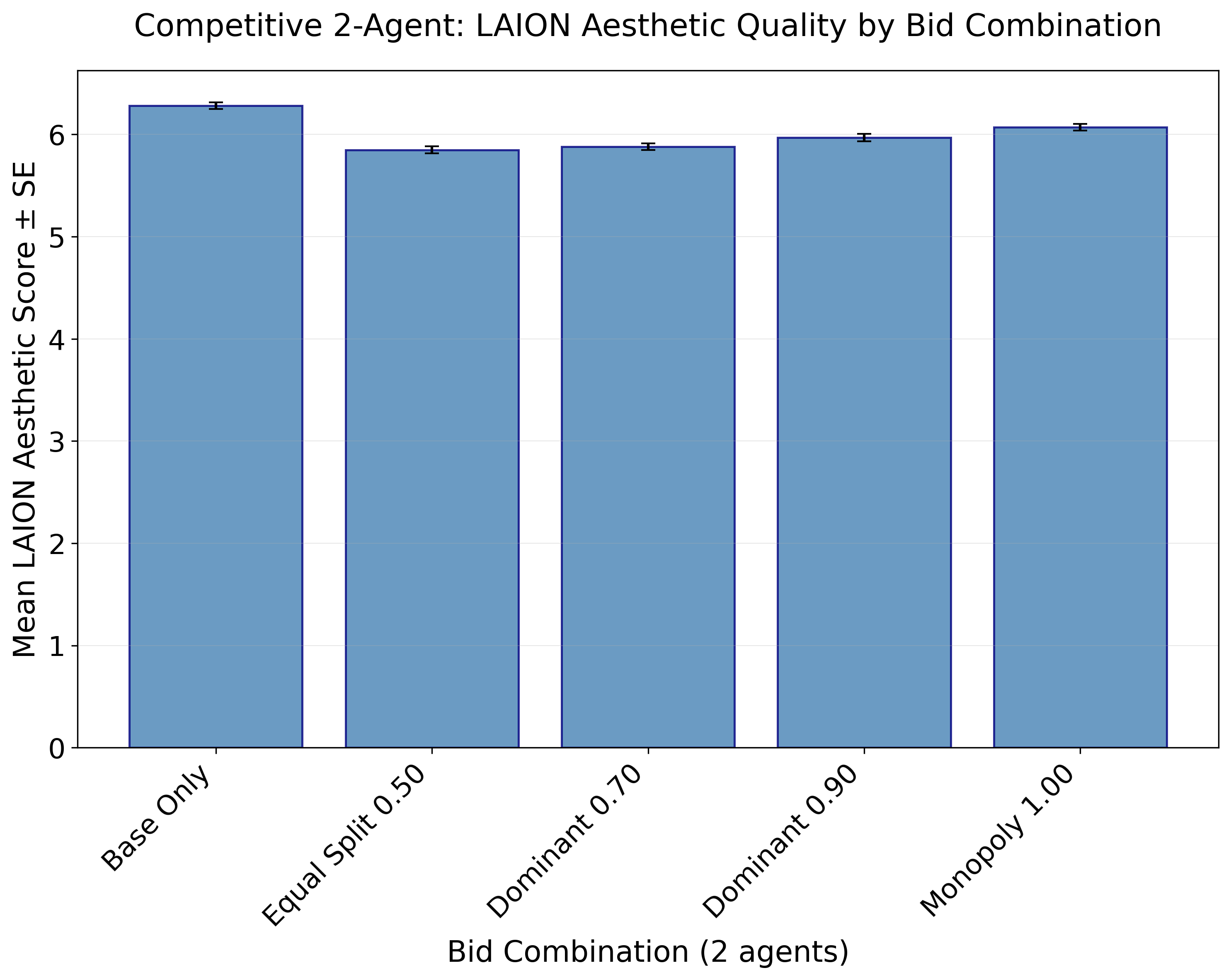}
        \caption{LAION image quality scores}
        \label{fig:laion_2agent_competitive}
    \end{subfigure}
    \caption{Competitive 2-agent bidding results.}
    \label{fig:2agent_competitive_results}
\end{figure}

\begin{figure}[H]
    \centering
    \begin{subfigure}[t]{0.45\linewidth}
        \centering
        \includegraphics[width=\linewidth]{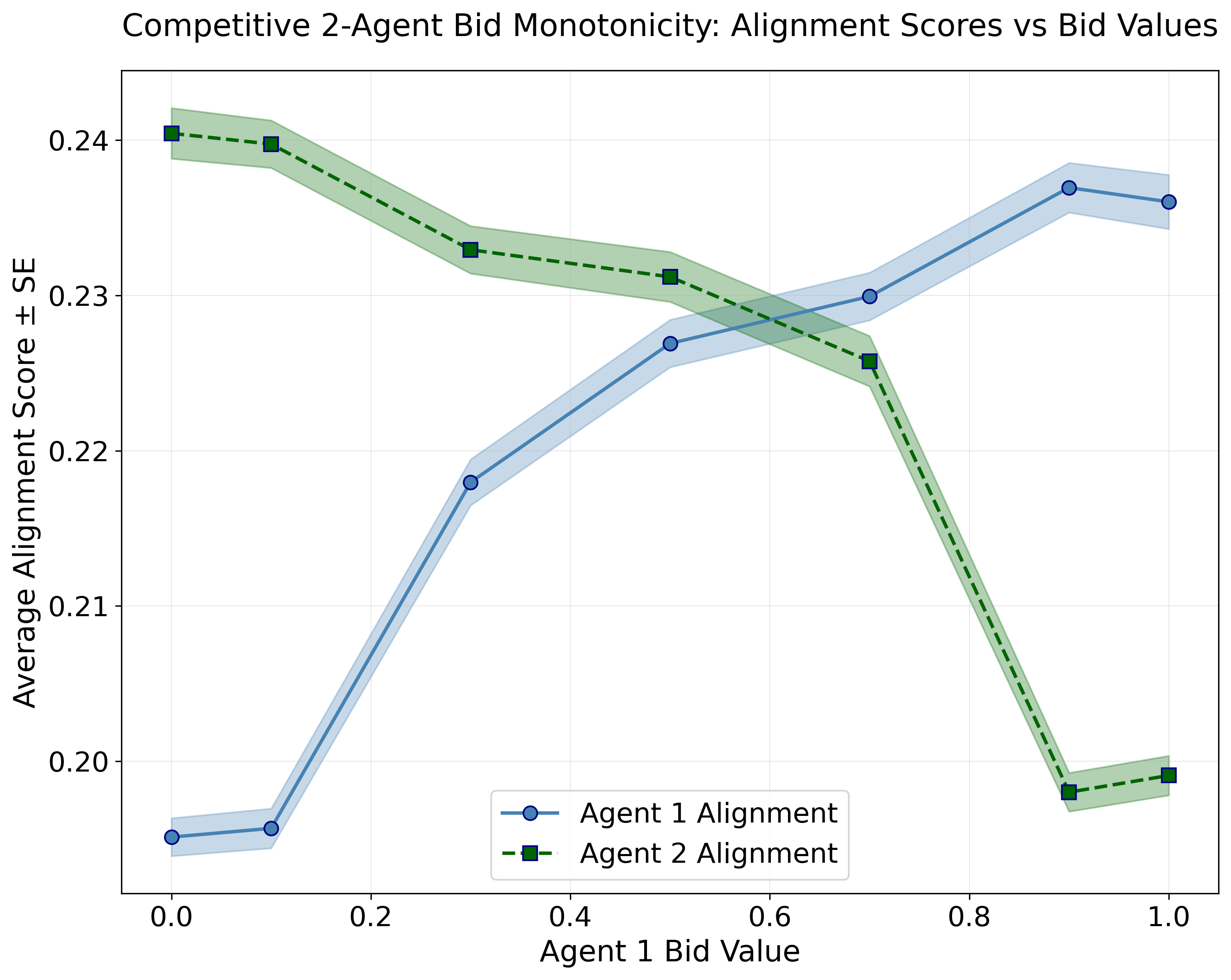}
        \caption{Approximate bid monotonicity.}
        \label{fig:bid_monotonicity_competitive}
    \end{subfigure}
    \hspace{1cm}
    \begin{subfigure}[t]{0.35\linewidth}
        \centering
        \includegraphics[width=\linewidth]{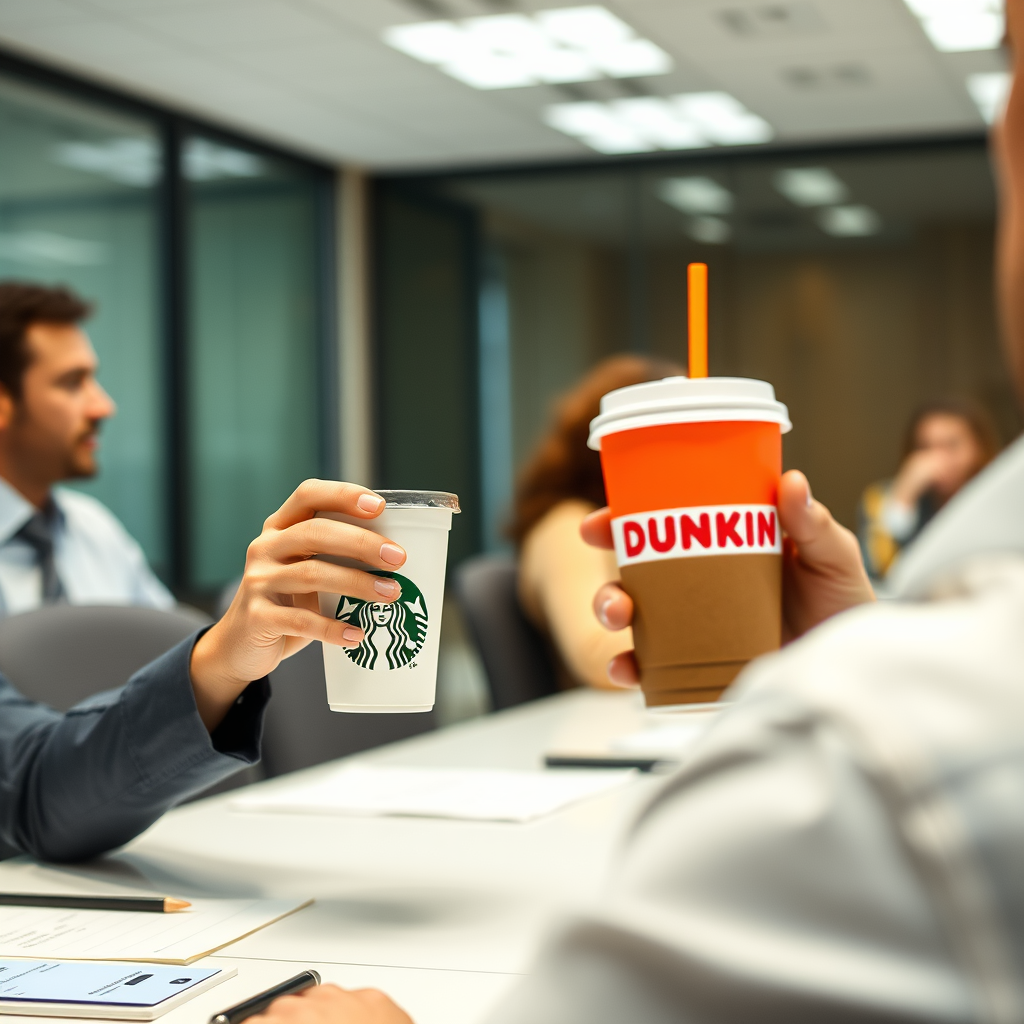}
        \caption{Example bidding scenario with Dunkin and Starbucks (0.5/0.5).}
        \label{fig:example_competitive}
    \end{subfigure}
    \caption{Allocation analysis for competitive 2-agent scenario.}
    \label{fig:competitive_combined}
\end{figure}

Our allocation rule still achieves welfare gains over the single-winner baseline of up to 3\%, though the magnitude is reduced compared to when the prompts are complementary; see Figure~\ref{fig:welfare_2agent_competitive}. Average alignment scores between the image chosen by the allocation rule and agents' prompts are approximately increasing with their bids, demonstrating that the allocation rule is approximately bid monotone;  see Figure \ref{fig:bid_monotonicity_competitive}. The images consistently achieve LAION scores above 5.75, indicating high image quality no matter the bid combinations; see Figure \ref{fig:laion_2agent_competitive}. 

These competitive bidding results address some of the most challenging real-world advertising scenarios. If advertisers want to avoid being shown next to their competitors, as in Figure~\ref{fig:example_competitive}, they can add additional instructions in their prompts that would reduce the welfare of such images, and therefore the propensity that such images are chosen by the allocation rule.





\section{Discussion}

\paragraph{High-Level Takeaway}
This work makes the argument that using Generative AI we are able to ``increase abundance,'' i.e.~create opportunities to improve welfare. We explore this in the context of banner ad auctions, showing that mechanism design theory can be combined with diffusion model practice to create mechanisms that are approximately truthful and attain higher welfare compared to standard single-winner mechanisms, by allowing advertisers to mix their advertising content in the image selected by the mechanism.

\paragraph{Applications} The framework we present offers strong potential for real-world applications, particularly in online advertising. Consider a search engine displaying banner ads on web pages: using our mechanism, it can generate visually compelling, contextually relevant images that reflect the preferences of multiple advertisers simultaneously. Our finding, that this generative mechanism yields higher welfare than single-winner mechanisms, provides a compelling incentive for platforms to adopt such methods. Advertisers benefit from having their branding or products meaningfully integrated into the generated content, with visibility and influence aligned with their bids. Beyond static images, our approach extends naturally to dynamic media. For example, a video platform could use our mechanism to embed branded elements, enabling multiple agents to bid for subtle, context-aware product placement. This opens new possibilities for non-intrusive, personalized advertising across multimedia formats.

\paragraph{Limitations}
A challenge for our setting and our approach arises from the fact that we are working with a rich allocation space, namely all natural images, which has no analytical description. Thus, we cannot optimize over it directly and resort to Monte Carlo sampling from a diffusion model to obtain samples in the high-weflare region of the allocation space.  The inherent probabilistic nature of denoising diffusion models, coupled with idiosyncrasy in how well text prompts are interpreted by the diffusion model, means that we can only hope to approximately optimize welfare for a given set of bids and prompts. This implies that truthfulness will only be approximate and some welfare will also be left on the table. While increasing the number of samples in the allocation rule is found to improve welfare and reduces sampling variance, improving the incentives, this comes at the cost of increased computational resources and generation time. There is thus a trade-off between economic and computational efficiency.

\paragraph{Future work.}
Our work is a first step at using diffusion models to improve efficiency in mechanism design settings, such as banner ad auctions, that can benefit from expanding the allocation space. There are several exciting directions to pursue next, including studying and controlling the impact of prompt manipulation by the bidders, evaluating the effect of directives in the bidder prompts, such as avoiding certain content or co-existence with competitors,  designing more sophisticated bidding languages that go beyond a single prompt and value, exploring revenue-maximizing versions of the mechanism, and adapting this mechanism to other  modalities like video.

\section{Conclusion}

Our work presents compelling evidence that diffusion models can be leveraged to increase abundance in mechanism design settings. We show how they can be used to express complex preferences and how they can be combined with mechanism design principles to give mechanisms that improve welfare compared to standard baselines and attain approximate incentive compatibility. We apply our approach to banner ad auctions, where diffusion models are leveraged to select images that aggregate the preferences of bidders according to the strengths of their bids. To this goal, we propose a score composition technique that enables prompt- and bid-dependent control over the content sampled by the diffusion model, in conjunction with payment rules that result in approximate incentive compatibility. Our comprehensive empirical analysis, conducted on a diverse dataset, comprising various prompt and bid combinations and considering up to 20-agent settings, provides compelling evidence for our mechanism's effectiveness, demonstrating improved social welfare over single-winner baselines and good incentive properties. Our research adds to the growing literature on the interface of Generative AI and Mechanism Design, and lays the groundwork for future investigations of how diffusion models and mechanism design can be combined to improve upon the properties of our mechanism as well as to extend it to other mechanism design objectives and other allocation modalities. 

%
%
%
%
%

\bibliographystyle{ACM-Reference-Format}
\bibliography{sample-bibliography}
\newpage
\appendix

\section{Prompts}
\label{app:prompts}

We prompted Claude to systematically generate base scenes spanning diverse everyday settings, each paired with  real-world brand prompts drawn from different product categories (food/beverage, tech, fashion, automotive, energy).

\vspace{-0.5\baselineskip}
\begin{table}[H]
\captionsetup{skip=2pt}
\caption{Example prompt configurations across $20$ three-agent scenarios}
\centering
\footnotesize
\begin{tabular}{>{\raggedright\arraybackslash}p{3.5cm} 
                >{\raggedright\arraybackslash}p{2.9cm} 
                >{\raggedright\arraybackslash}p{2.9cm}
                >{\raggedright\arraybackslash}p{2.9cm}}
\toprule
\textbf{Base Prompt} & \textbf{Agent 1 Prompt} & \textbf{Agent 2 Prompt} & \textbf{Agent 3 Prompt} \\
\midrule
Two friends chatting over coffee at a cafe & Starbucks mug & Apple MacBook Air & New York Times newspaper \\ \hline
People enjoying a sunny day at the beach & Wilson volleyball & Red Bull cooler & Banana Boat sunscreen \\ \hline
A person relaxing on a living-room sofa & Lay's potato chips & M\&M's candy & Netflix show on TV \\ \hline
A hiker resting on a mountain trail & North Face backpack & Gatorade bottle & Garmin watch \\ \hline
A professional working at a desk in a bright office & Dell monitor & Deer Park water bottle & Moleskine notebook \\ \hline
A woman with a tray of food at a cafeteria & McDonald's fries & Pepsi can & Heinz ketchup packet \\ \hline
A person walking down a city street & Louis Vuitton handbag & H\&M storefront & Nike sneakers \\ \hline
Friends gathered for a movie night at home & Doritos chips & Coca-Cola cans & Amazon Fire TV remote \\ \hline
A car driving along a scenic coastal highway & Tesla car & ExxonMobil sign & Goodyear tires \\ \hline
Skiers waiting in a ski-lift line & Patagonia jacket & GoPro camera & Beats wireless headphones \\ \hline
A man waiting at a bus stop & Adidas backpack & Samsung Galaxy phone & Sony headphones \\ \hline
A student studying in a university library & MacBook Pro laptop & Red Bull energy drink & Coca-Cola can \\ \hline
A chef cooking in a restaurant kitchen & KitchenAid stand mixer & Coca-Cola glass bottle & Heinz ketchup bottle \\ \hline
A couple shopping at a farmers market & Whole Foods tote bag & iPhone camera & Patagonia fleece jacket \\ \hline
A teenager gaming in their bedroom & PlayStation 5 controller & Monster Energy drink & Razer gaming headset \\ \hline
A businesswoman in an airport lounge & Louis Vuitton luggage & Starbucks coffee cup & iPad Pro tablet \\ \hline
A mechanic working in an auto garage & Ford truck & DeWalt power drill & Mobil 1 motor oil \\ \hline
A barista working in a trendy coffee shop & Starbucks coffee machine & Spotify playlist screen & Levi's denim apron \\ \hline
A fitness trainer at a modern gym & Under Armour workout shirt & Beats wireless headphones & Hydroflask water bottle \\ \hline
A doctor in a hospital break room & MacBook Air laptop & Starbucks coffee cup & Samsung tablet \\
\bottomrule
\end{tabular}
\label{tab:prompts_3_agents}
\end{table}

\begin{table}[H]
\centering
\captionsetup{skip=2pt}
\caption{Number of Prompts per Agent Configuration}
\begin{tabular}{cc}
\hline
\textbf{Number of Agents} & \textbf{Number of Prompts} \\
\hline
2 & 200 \\
3 & 133 \\
5 & 85 \\
10 & 44 \\
20 & 25 \\
\hline
\end{tabular}
\label{tab:prompts_per_config}
\end{table}

\renewcommand{\arraystretch}{1.0} 

\section{Generalized Score Composition}
\label{app:generalized_score}

The score composition can be extended to $n$ agents ($s^{(1:n)}_t(x)$) by recursively defining it in terms of score compositions for $n-1$ agents, $n-2$ agents, etc. Assume, without loss of generality, that agents $1$ to $n$ submit bids in decreasing value: $b_1 \geq b_2 \geq \ldots \geq b_n$. Denote $b_j^{(0)} = b_j$. For recursive step $i$, take $b_j^{(i)} = \frac{b_j^{(i-1)}}{\sum_{k=1}^{n-i+1} b_k^{(i-1)}}$ to be the normalized bids and $w^{(i)} = \left({n-i+1 \over n-i}\sum_{j=1}^{n-i} b_j^{(i)} - 1\right)$ to be the normalized weights. Denote by $c_{1:i} = \{c_1, c_2, \ldots, c_i\}$  the concatenated prompt of agents $1$ through $i$, and take $s^{(1:1)}_t(x) := s(x|c, c_1)$ to be the conditional score conditioning on the dominant agent's prompt (and the base prompt if any).  We then define the composite score $s^{(1:n)}_t(x)$ of all $n$ agents by separating out the least dominant agent, $n$, and computing the score function for the remaining $n-1$ agents recursively as follows:
\begin{equation}
s^{(1:n)}_t(x) = (1-w^{(1)}) \, s(x|c, c_{1:n}) + w^{(1)} \, s^{(1:n-1)}_t(x)
\end{equation}
\begin{equation}
s^{(1:i)}_t(x) = (1-w^{(n-i+1)}) \, s(x|c, c_{1:i}) + w^{(n-i+1)} \, s^{(1:i-1)}_t(x)
\end{equation}
We can also derive a closed-form expression for $s^{(1:n)}_t(x)$.
\begin{equation}
s^{(1:n)}_t(x) = \prod_{i=1}^{n-1} w^{(i)} \, s(x|c,c_1) + \sum_{i=2}^{n}\left[\left(\prod_{j=1}^{n-i} w^{(j)}\right)(1-w^{(n-i+1)})\, s(x|c,c_{1:i})\right]
\end{equation}
\newpage
\section{Linear Aggregation}
\label{app:linear_aggregation}

\begin{figure}[H]
    \centering
    \includegraphics[width=1\linewidth]{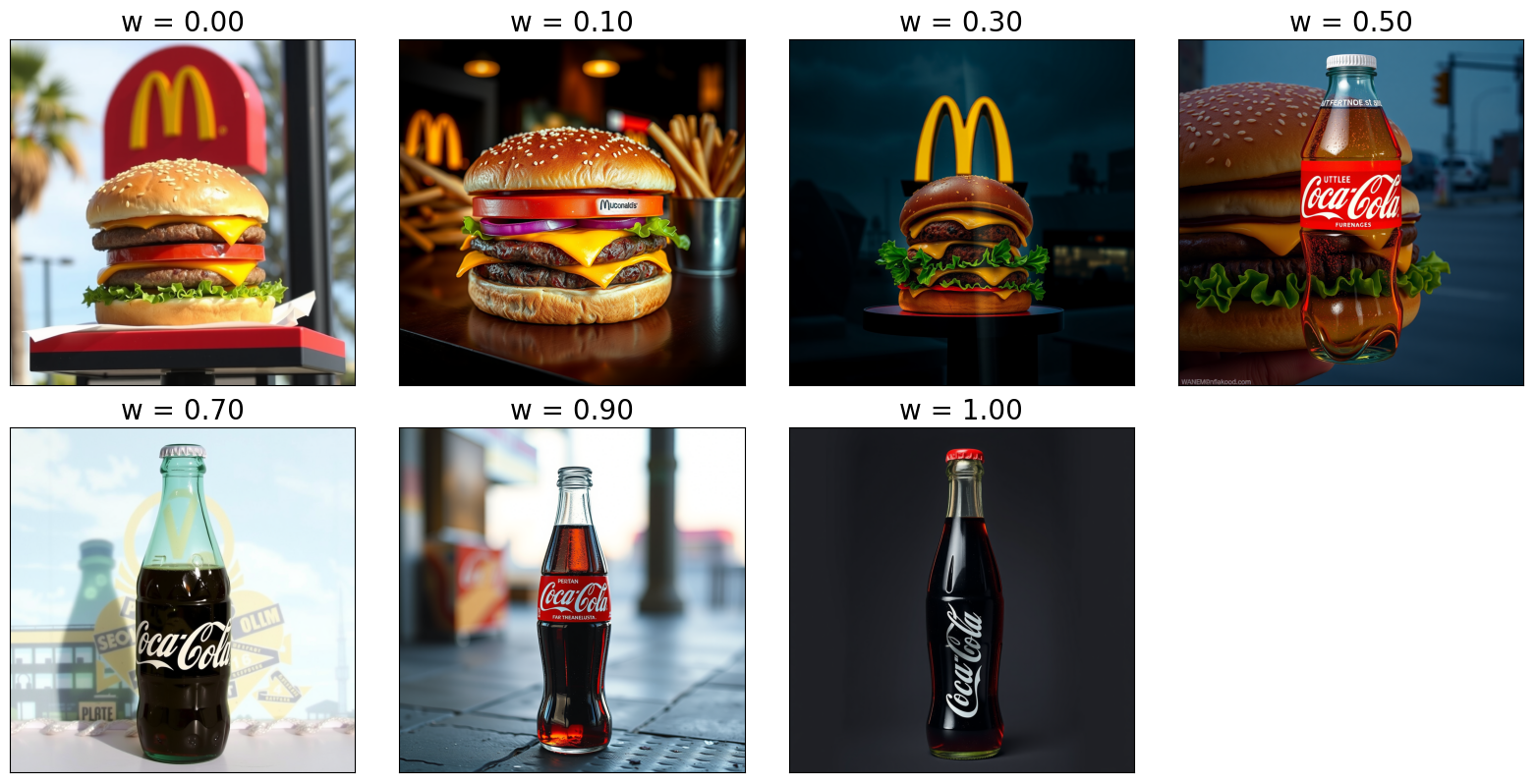}
    \caption{Linear aggregation function without the joint score fails to produce compositional images.}
    \label{fig:linear_aggregation}
\end{figure}

We had first considered the linear aggregation function. \begin{equation}
    s^{(1,2)}_t(x) = (1-w^{(1)})s_t(x|c,c_1) + (w^{(1)})s_t(x|c,c_2)
\end{equation}
However, we observed that it did not yield compositional images. Instead, the objects were faded on top of one another, leading to visually incoherent outcomes (Figure \ref{fig:linear_aggregation}). 

\section{Comprehensive Empirical Evaluation}
\subsection{Welfare Improvement}
\label{app:welfare_all_agents}
\begin{figure}[H]
    \centering
    \begin{subfigure}[b]{0.44\linewidth}
        \centering
        \includegraphics[width=\linewidth]{images_new/welfare_clip_over_bids_2_agents.png}
        \caption{2 agents, 200 prompts}
        \label{fig:welfare_samples_2agents}
    \end{subfigure}
    \hfill
    \begin{subfigure}[b]{0.44\linewidth}
        \centering
        \includegraphics[width=\linewidth]{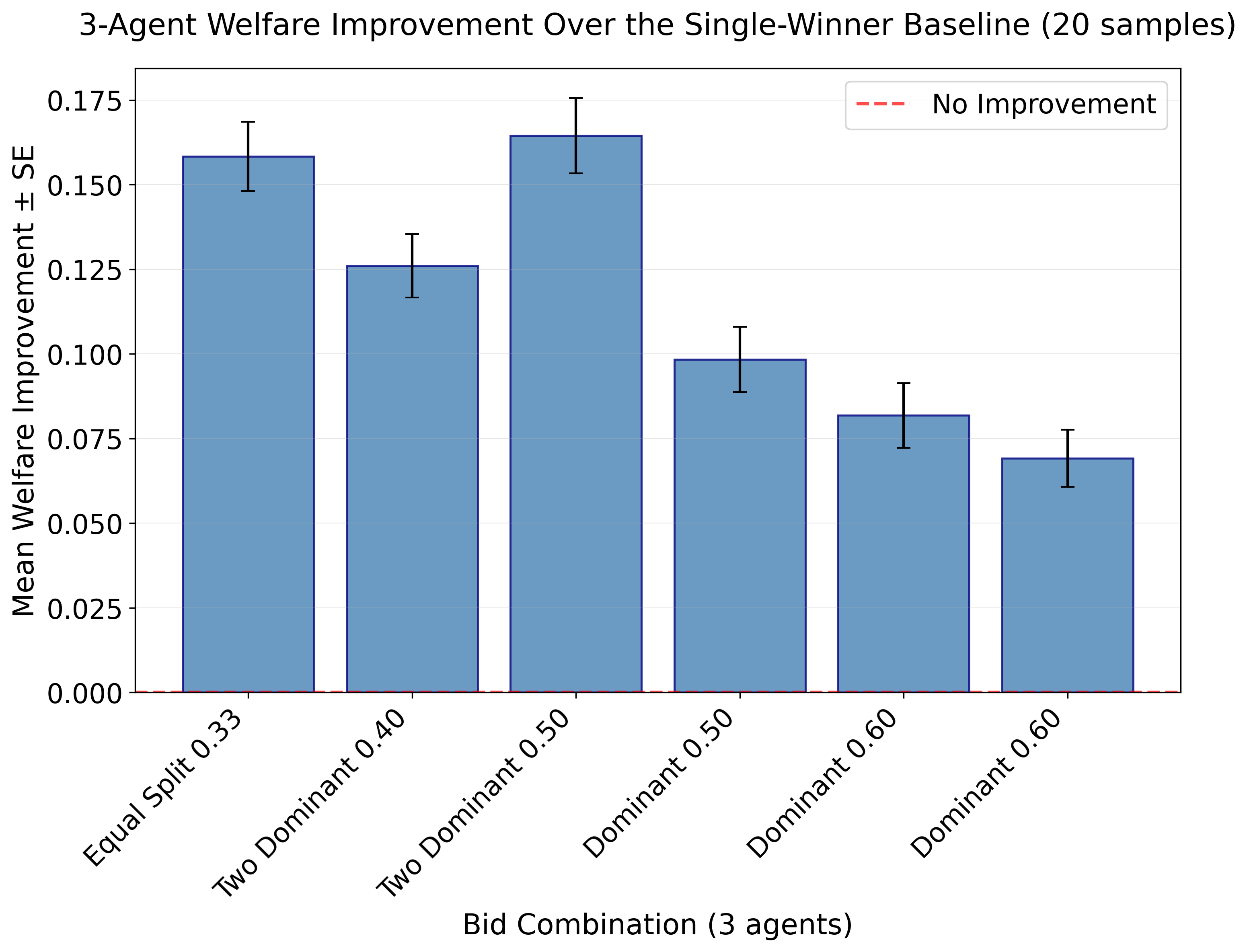}
        \caption{3 agents, 133 prompts}
        \label{fig:welfare_samples_3agents}
    \end{subfigure}
    
    \vspace{0.3cm}
    
    \begin{subfigure}[b]{0.44\linewidth}
        \centering
        \includegraphics[width=\linewidth]{images_new/welfare_clip_over_bids_5_agents.png}
        \caption{5 agents, 85 prompts}
        \label{fig:welfare_samples_5agents}
    \end{subfigure}
    \hfill
    \begin{subfigure}[b]{0.44\linewidth}
        \centering
        \includegraphics[width=\linewidth]{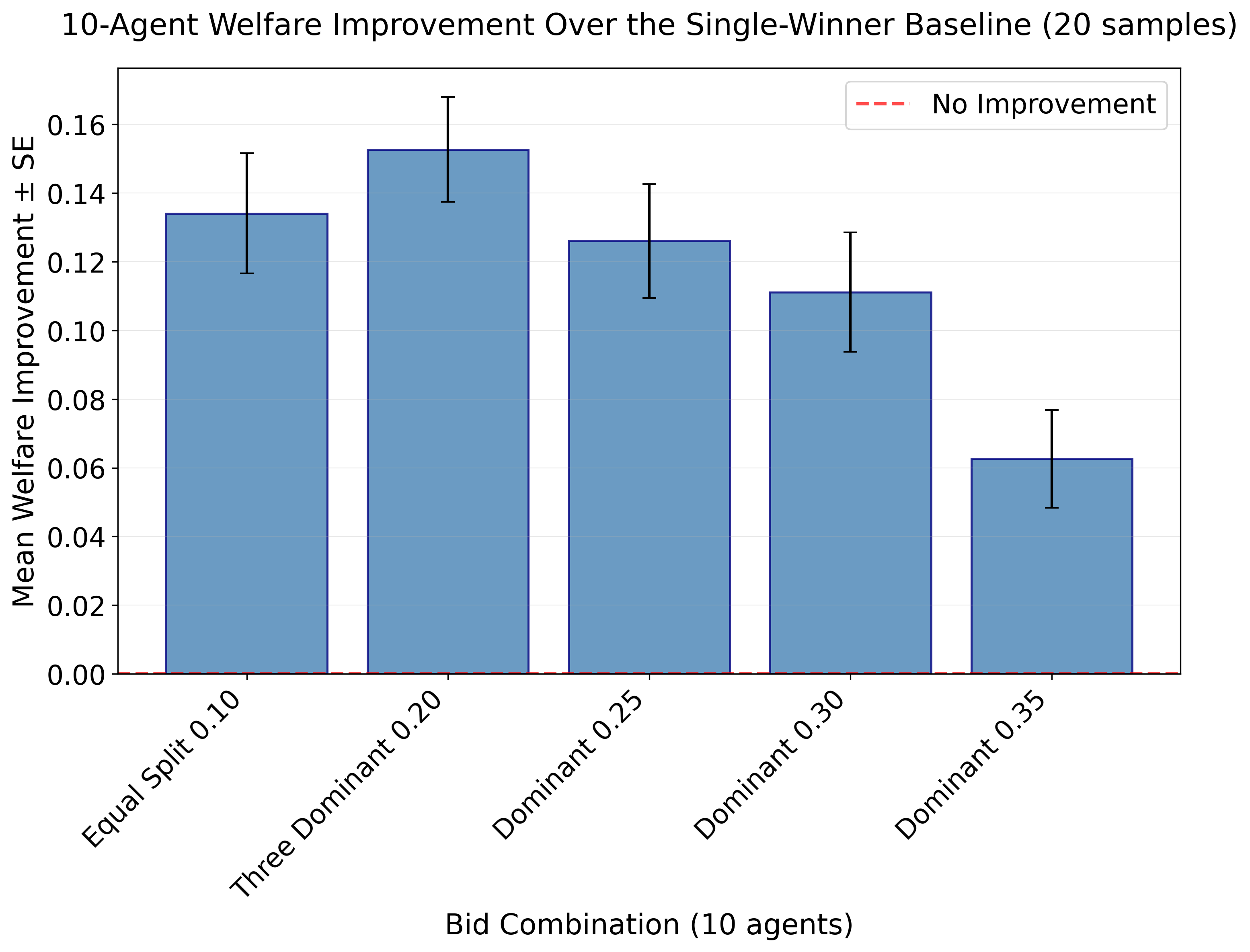}
        \caption{10 agents, 44 prompts}
        \label{fig:welfare_samples_10agents}
    \end{subfigure}
    
    \vspace{0.3cm}
    
    \begin{subfigure}[b]{0.44\linewidth}
        \centering
        \includegraphics[width=\linewidth]{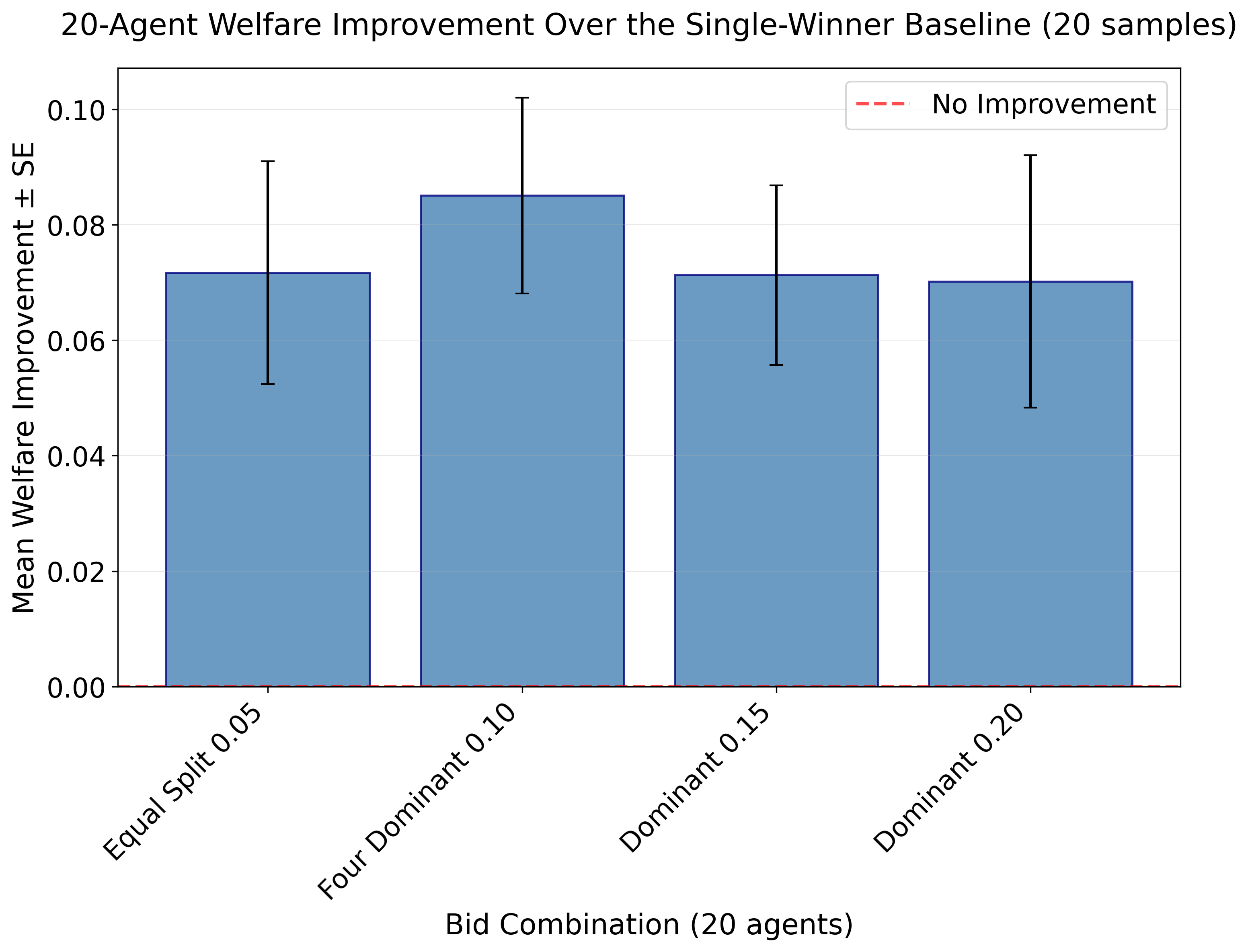}
        \caption{20 agents, 25 prompts}
        \label{fig:welfare_samples_20agents}
    \end{subfigure}
    
    \caption{Welfare improvement over single-winner baseline from 2 to 20 agents with varying bid combinations, exhibiting up to 23\% increase in the 5 agent setting.}
    \label{fig:welfare_samples_all_agents}
\end{figure}

\subsection{LAION Image Quality}
\label{app:laion}
\begin{figure}[H]
    \centering
    \begin{subfigure}[b]{0.44\linewidth}
        \centering
        \includegraphics[width=\linewidth]{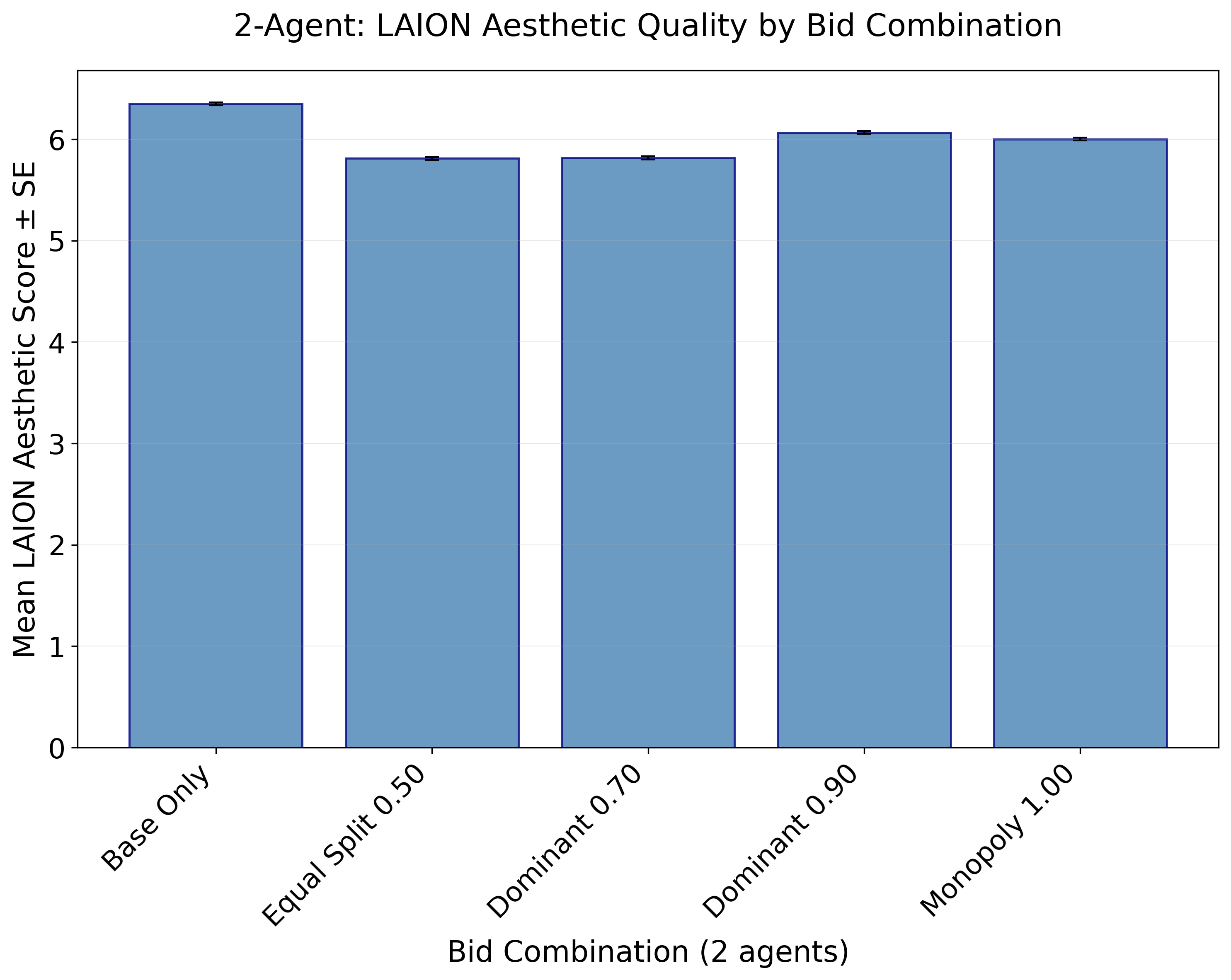}
        \caption{2 agents}
        \label{fig:laion_quality_2agents}
    \end{subfigure}
    \hfill
    \begin{subfigure}[b]{0.44\linewidth}
        \centering
        \includegraphics[width=\linewidth]{images_new/laion_quality_bar_3_agents.png}
        \caption{3 agents}
        \label{fig:laion_quality_3agents}
    \end{subfigure}
    
    \vspace{0.3cm}
    
    \begin{subfigure}[b]{0.44\linewidth}
        \centering
        \includegraphics[width=\linewidth]{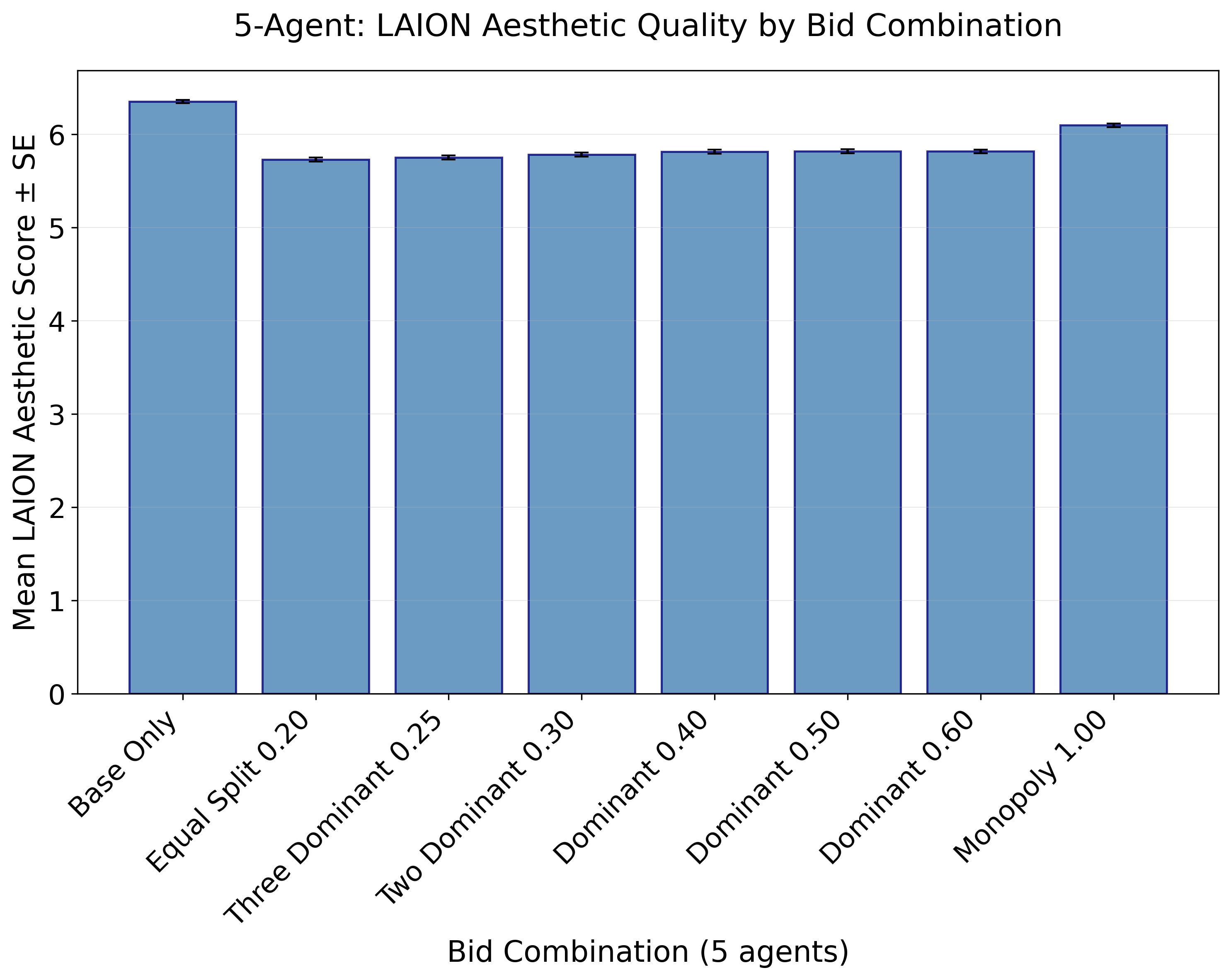}
        \caption{5 agents}
        \label{fig:laion_quality_5agents}
    \end{subfigure}
    \hfill
    \begin{subfigure}[b]{0.44\linewidth}
        \centering
        \includegraphics[width=\linewidth]{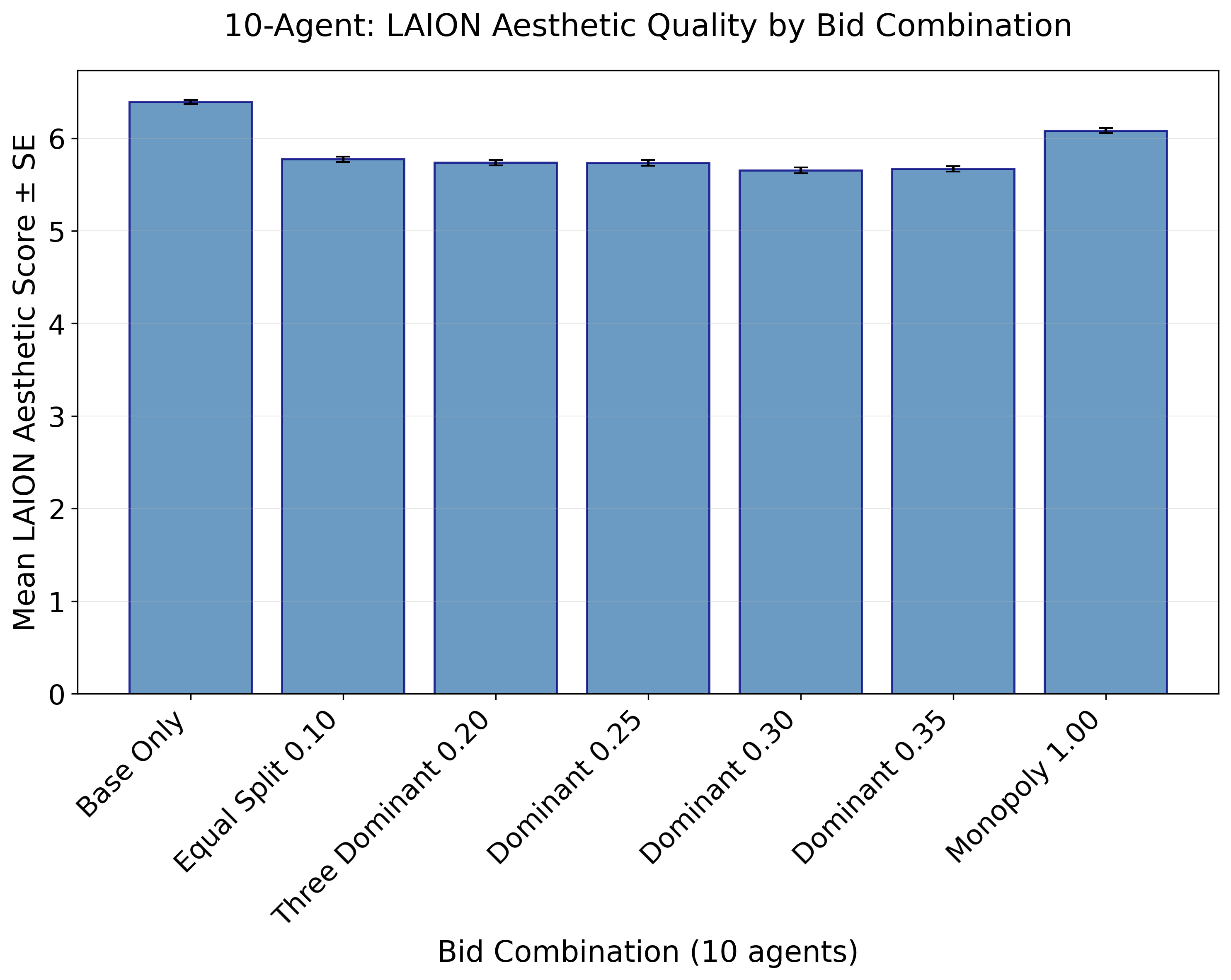}
        \caption{10 agents}
        \label{fig:laion_quality_10agents}
    \end{subfigure}
    
    \vspace{0.3cm}
    
    \begin{subfigure}[b]{0.44\linewidth}
        \centering
        \includegraphics[width=\linewidth]{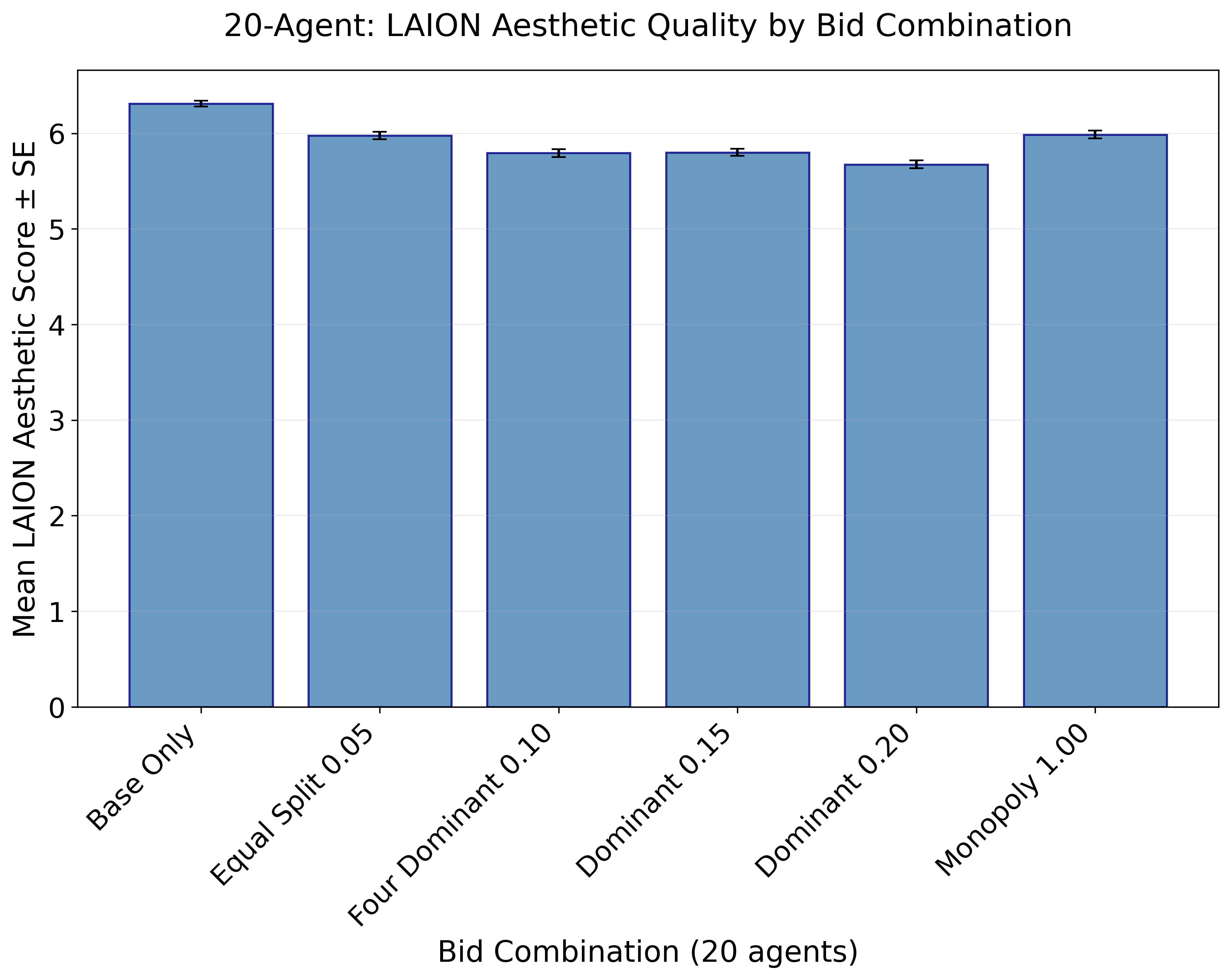}
        \caption{20 agents}
        \label{fig:laion_quality_20agents}
    \end{subfigure}
    
    \caption{LAION aesthetic quality scores for varying numbers of agents. Scores are shown for different bid configurations for the agents. Scores are computed across all the prompt configurations we consider in Appendix~\ref{app:prompts}. Scores consistently remain above 5.75 across, indicating high image quality is preserved regardless of the number of agents or bid configurations. }
    \label{fig:laion_quality_all_agents}
\end{figure}

\newpage
\subsection{Image Quality using CLIP Alignment}
\label{app:clip-alignment}

\begin{table}[htb]
\centering
\begin{minipage}{1\textwidth}
\centering
\caption{Image quality in the 2 agent setting according to CLIP alignment score. Computed across prompt configurations in Appendix~\ref{app:prompts} and various bid combinations.}
\label{tab:image_quality_2_agent}
\begin{tabular}{cccc}
\hline
Agent 1 & Agent 2 & Mean & SD \\
\hline
0.0 & 1.0 & 0.5081 & 0.0028 \\
0.1 & 0.9 & 0.5084 & 0.0030 \\
0.3 & 0.7 & 0.5083 & 0.0031 \\
0.5 & 0.5 & 0.5082 & 0.0030 \\
0.7 & 0.3 & 0.5085 & 0.0028 \\
0.9 & 0.1 & 0.5090 & 0.0027 \\
1.0 & 0.0 & 0.5089 & 0.0028 \\
\hline
\end{tabular}
\end{minipage}
\hfill
\vspace{1cm}
\begin{minipage}{1\textwidth}
\centering
\caption{Image quality in the 3 agent setting according to CLIP alignment score. Computed across prompt configurations in Appendix~\ref{app:prompts}  and various bid combinations.}
\label{tab:image_quality_3_agent}
\begin{tabular}{ccccc}
\hline
Agent 1 & Agent 2 & Agent 3 & Mean & SD \\
\hline
0.33 & 0.33 & 0.33 & 0.5084 & 0.0029 \\
0.40 & 0.40 & 0.20 & 0.5083 & 0.0030 \\
0.60 & 0.30 & 0.10 & 0.5082 & 0.0029 \\
0.60 & 0.20 & 0.20 & 0.5086 & 0.0029 \\
1.00 & 0.00 & 0.00 & 0.5088 & 0.0027 \\
\hline
\end{tabular}
\end{minipage}
\end{table}

\subsection{Monte Carlo Samples Ablation}
\label{app:mc_all_agents}
\begin{figure}[H]
    \centering
    \begin{subfigure}[b]{0.44\linewidth}
        \centering
        \includegraphics[width=\linewidth]{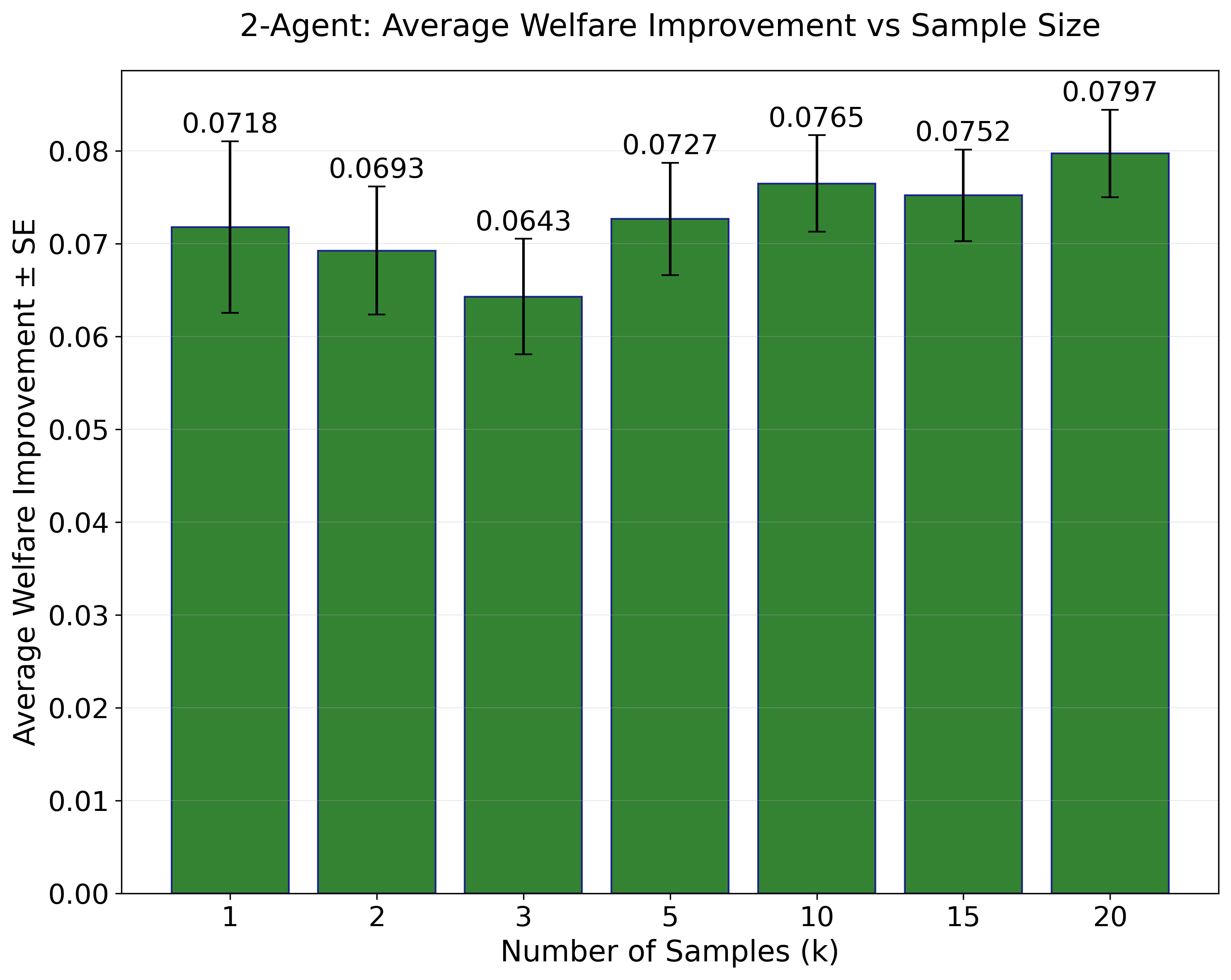}
        \caption{2 agents}
        \label{fig:welfare_k_2agents}
    \end{subfigure}
    \hfill
    \begin{subfigure}[b]{0.44\linewidth}
        \centering
        \includegraphics[width=\linewidth]{images_new/welfare_clip_over_k_3_agents.png}
        \caption{3 agents}
        \label{fig:welfare_k_3agents}
    \end{subfigure}
    
    \vspace{0.3cm}
    
    \begin{subfigure}[b]{0.44\linewidth}
        \centering
        \includegraphics[width=\linewidth]{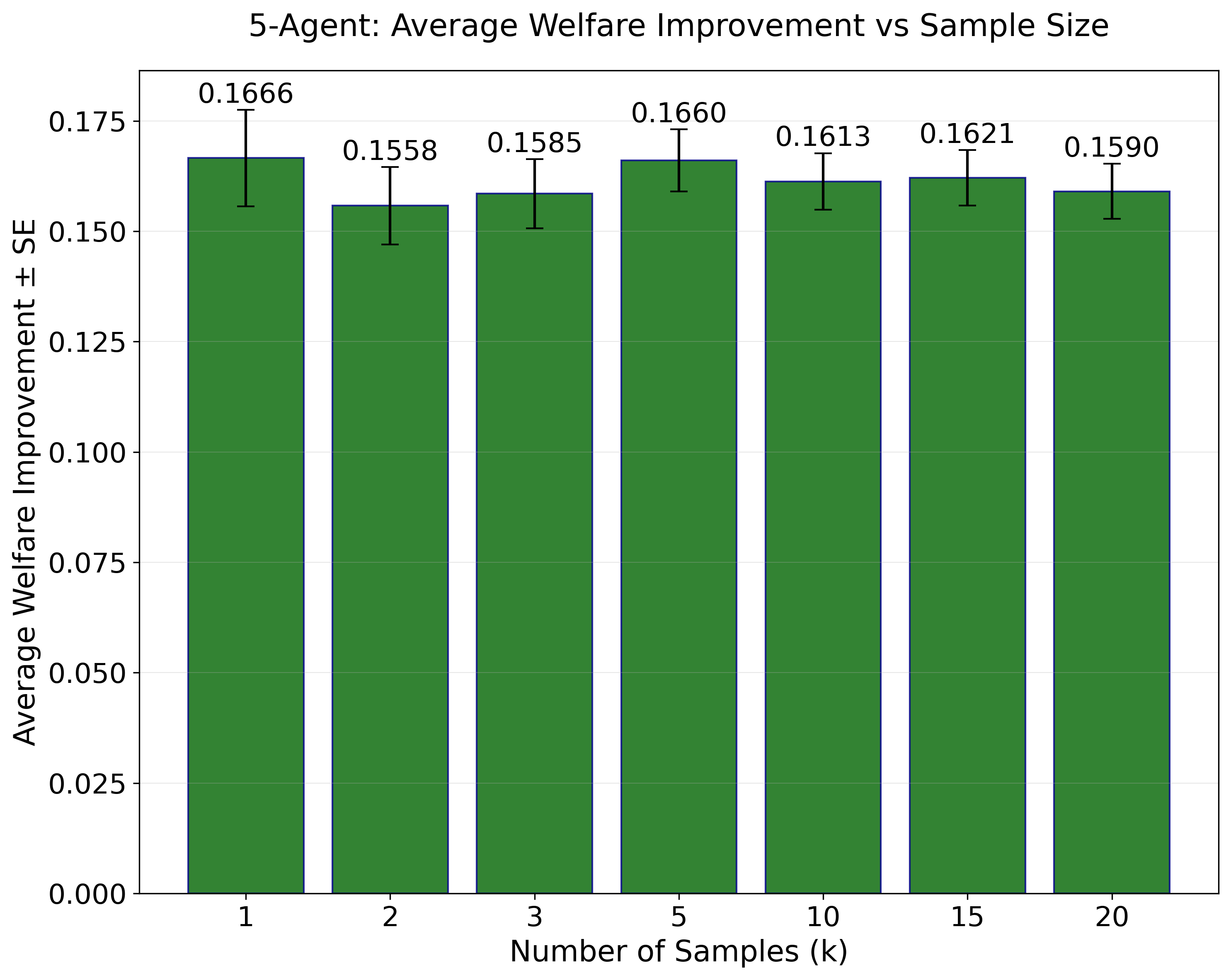}
        \caption{5 agents}
        \label{fig:welfare_k_5agents}
    \end{subfigure}
    \hfill
    \begin{subfigure}[b]{0.44\linewidth}
        \centering
        \includegraphics[width=\linewidth]{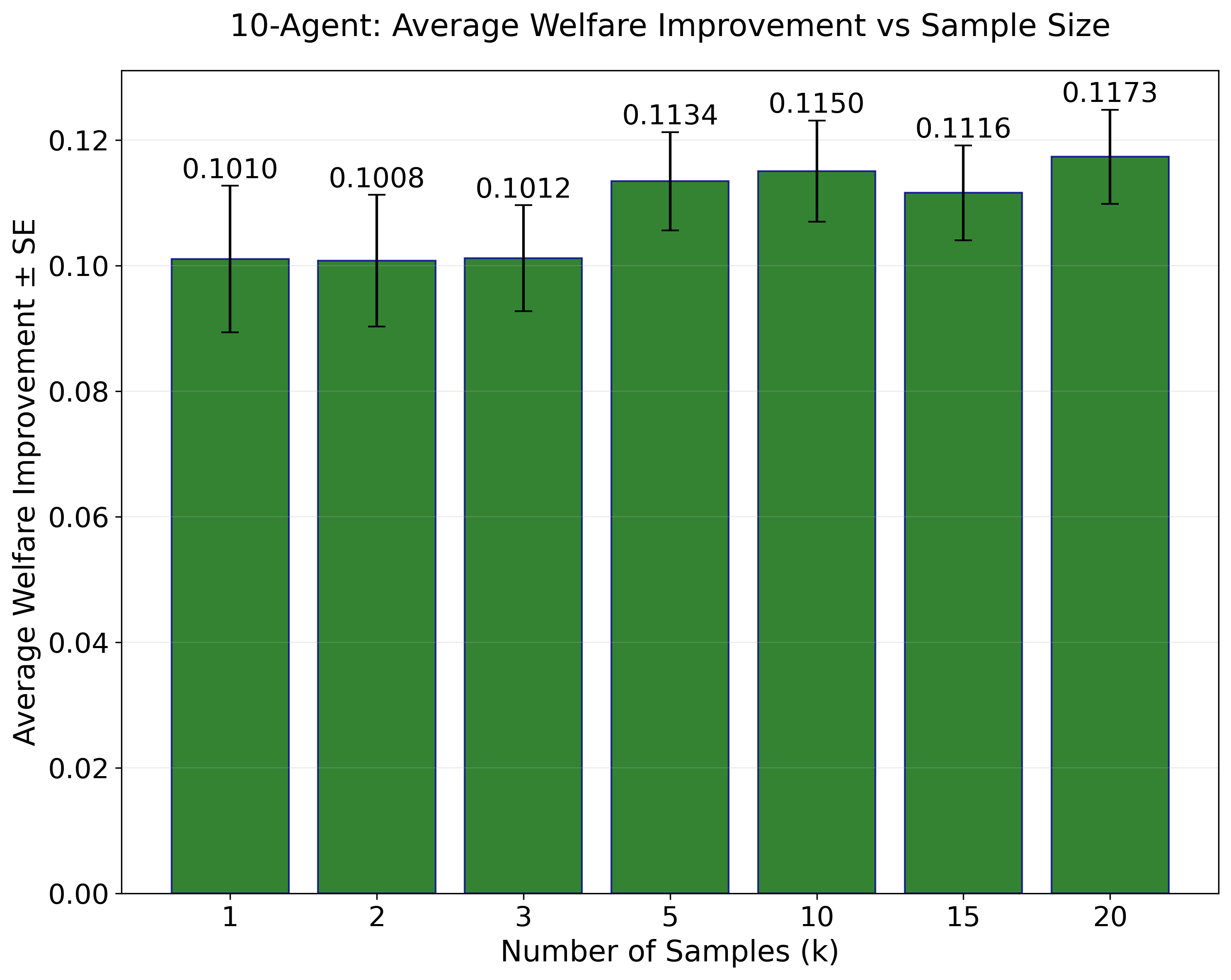}
        \caption{10 agents}
        \label{fig:welfare_k_10agents}
    \end{subfigure}
    
    \vspace{0.3cm}
    
    \begin{subfigure}[b]{0.44\linewidth}
        \centering
        \includegraphics[width=\linewidth]{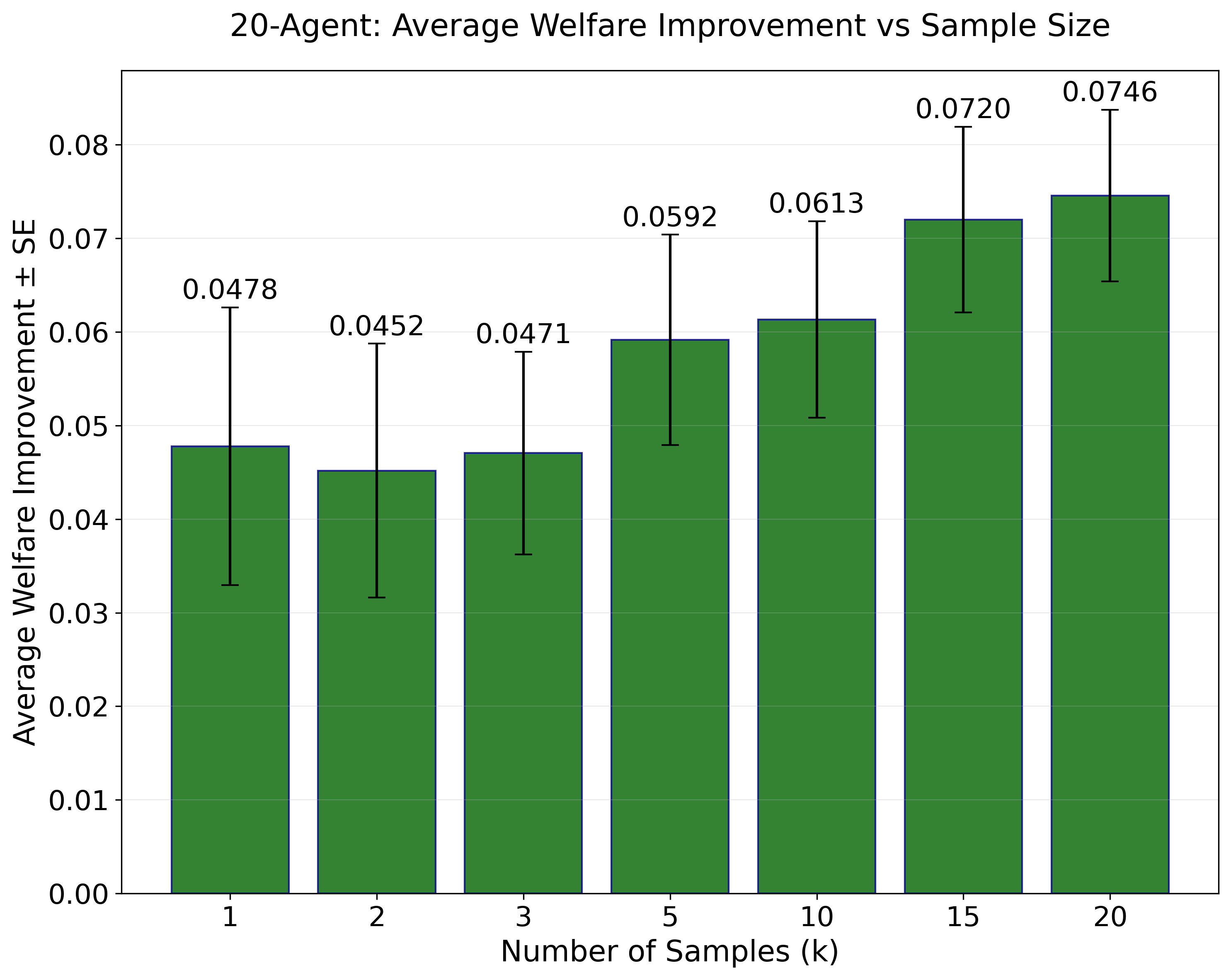}
        \caption{20 agents}
        \label{fig:welfare_k_20agents}
    \end{subfigure}
    
    \caption{Welfare improvement over the single-winner baseline as a function of the number $k$ of Monte Carlo samples in the allocation rule. For 2, 3, 5, 10 agents, welfare improvement plateaus around $k=5$ samples, where it reaches at least 90\% of the welfare improvement attained with $k=20$. For 20 agents, there is room for improvement beyond $k=5$, but $k=15$ attains at least 95\% of the welfare improvement at $k=20$. 
    }
    \label{fig:welfare_k_all_agents}
\end{figure}

\subsection{Competitive Bidding}
\label{app:competitive_bidding}
\begin{table}[H]
\caption{Prompt configurations across $20$ two-agent competitive scenarios}
\centering
\footnotesize
\begin{tabular}{>{\raggedright\arraybackslash}p{6cm} 
                >{\raggedright\arraybackslash}p{3cm} 
                >{\raggedright\arraybackslash}p{3cm}}
\toprule
\textbf{Base Prompt} & \textbf{Agent 1 Prompt} & \textbf{Agent 2 Prompt} \\
\midrule
A person drinking from a beverage can while working at a desk & Coca-Cola can & Pepsi can \\ \hline
Someone wearing athletic shoes while jogging in a park & Nike running shoes & Adidas running shoes \\ \hline
A person wearing a smartwatch while exercising at the gym & Apple Watch & Samsung Galaxy Watch \\ \hline
Someone drinking from a water bottle during a yoga class & Hydro Flask water bottle & Yeti water bottle \\ \hline
A driver filling up their car at a gas station & Shell gas station & Chevron gas station \\ \hline
A customer holding a shopping bag from a clothing store & Zara shopping bag & H\&M shopping bag \\ \hline
A person wearing wireless earbuds while commuting & Apple AirPods & Samsung Galaxy Buds \\ \hline
Someone drinking an energy drink at a skateboard park & Red Bull can & Monster Energy can \\ \hline
A person carrying a backpack while hiking on a trail & North Face backpack & Patagonia backpack \\ \hline
Someone using a smartphone at a coffee shop & iPhone smartphone & Samsung Galaxy phone \\ \hline
A person wearing a baseball cap at a sports game & New York Yankees cap & Boston Red Sox cap \\ \hline
Someone drinking from a coffee cup in an office meeting & Starbucks coffee cup & Dunkin coffee cup \\ \hline
A customer carrying a fast food bag on a city street & McDonald's bag & Burger King bag \\ \hline
A person wearing running shorts during a marathon & Nike running shorts & Adidas running shorts \\ \hline
Someone eating chips while watching TV at home & Lay's chips bag & Doritos bag \\ \hline
A person wearing a winter jacket in a snowy setting & Canada Goose jacket & North Face jacket \\ \hline
Someone drinking orange juice at breakfast & Tropicana orange juice & Minute Maid orange juice \\ \hline
A person wearing a fitness tracker at a cycling class & Fitbit tracker & Garmin tracker \\ \hline
Someone holding a pizza box at a house party & Domino's pizza box & Pizza Hut box \\ \hline
A person wearing a hoodie at a college campus & Champion hoodie & Nike hoodie \\
\bottomrule
\end{tabular}
\label{tab:prompts_2_agents_competitive}
\end{table}
\renewcommand{\arraystretch}{1.0}

\newpage

\section{Bid Combination Values}
\label{app:bid-dist}

\begin{table}[H]
\centering
\captionsetup{skip=2pt}
\caption{Bid Combinations for 2 Bidders}
\begin{tabular}{ll}
\hline
\textbf{Combination Name} & \textbf{Bid Values} \\
\hline
Equal Split 0.50 & $(0.50, 0.50)$ \\
Dominant 0.70 & $(0.30, 0.70)$ \\
Dominant 0.90 & $(0.10, 0.90)$ \\
\hline
\end{tabular}
\label{tab:bids_2}
\end{table}

\begin{table}[H]
\centering
\captionsetup{skip=2pt}
\caption{Bid Combinations for 3 Bidders}
\begin{tabular}{ll}
\hline
\textbf{Combination Name} & \textbf{Bid Values} \\
\hline
Equal Split 0.33 & $(0.33, 0.33, 0.33)$ \\
Two Dominant 0.40 & $(0.40, 0.40, 0.20)$ \\
Two Dominant 0.50 & $(0.50, 0.50, 0.00)$ \\
Dominant 0.50 & $(0.50, 0.30, 0.20)$ \\
Dominant 0.60 & $(0.60, 0.30, 0.10)$ \\
Dominant 0.60 & $(0.60, 0.20, 0.20)$ \\
\hline
\end{tabular}
\label{tab:bids_3}
\end{table}

\begin{table}[H]
\centering
\captionsetup{skip=2pt}
\caption{Bid Combinations for 5 Bidders}
\begin{tabular}{ll}
\hline
\textbf{Combination Name} & \textbf{Bid Values} \\
\hline
Equal Split 0.20 & $(0.20, 0.20, 0.20, 0.20, 0.20)$ \\
Three Dominant 0.25 & $(0.25, 0.25, 0.25, 0.15, 0.10)$ \\
Two Dominant 0.30 & $(0.30, 0.30, 0.20, 0.10, 0.10)$ \\
Dominant 0.40 & $(0.40, 0.30, 0.20, 0.05, 0.05)$ \\
Dominant 0.50 & $(0.50, 0.25, 0.15, 0.07, 0.03)$ \\
Dominant 0.60 & $(0.60, 0.20, 0.10, 0.05, 0.05)$ \\
\hline
\end{tabular}
\label{tab:bids_5}
\end{table}

\begin{table}[H]
\centering
\captionsetup{skip=2pt}
\caption{Bid Combinations for 10 Bidders}
\begin{tabular}{ll}
\hline
\textbf{Combination Name} & \textbf{Bid Values} \\
\hline
Equal Split 0.10 & $(0.10, 0.10, 0.10, 0.10, 0.10, 0.10, 0.10, 0.10, 0.10, 0.10)$ \\
Three Dominant 0.20 & $(0.20, 0.20, 0.20, 0.15, 0.10, 0.05, 0.05, 0.03, 0.01, 0.01)$ \\
Dominant 0.25 & $(0.25, 0.20, 0.15, 0.12, 0.10, 0.08, 0.05, 0.03, 0.01, 0.01)$ \\
Dominant 0.30 & $(0.30, 0.20, 0.15, 0.12, 0.10, 0.07, 0.04, 0.01, 0.00, 0.01)$ \\
Dominant 0.35 & $(0.35, 0.15, 0.12, 0.10, 0.08, 0.07, 0.05, 0.04, 0.02, 0.02)$ \\
\hline
\end{tabular}
\label{tab:bids_10}
\end{table}

\begin{table}[H]
\centering
\captionsetup{skip=2pt}
\caption{Bid Combinations for 20 Bidders}
\begin{tabular}{ll}
\hline
\textbf{Combination Name} & \textbf{Bid Values} \\
\hline
Equal Split 0.05 & $(0.05, 0.05, 0.05, 0.05, 0.05, 0.05, 0.05, 0.05, 0.05, 0.05,$ \\
 & $0.05, 0.05, 0.05, 0.05, 0.05, 0.05, 0.05, 0.05, 0.05, 0.05)$ \\
Four Dominant 0.10 & $(0.10, 0.10, 0.10, 0.10, 0.08, 0.08, 0.08, 0.07, 0.07, 0.07,$ \\
 & $0.05, 0.05, 0.05, 0.02, 0.02, 0.02, 0.01, 0.01, 0.01, 0.01)$ \\
Dominant 0.15 & $(0.15, 0.12, 0.10, 0.08, 0.07, 0.06, 0.05, 0.05, 0.04, 0.04,$ \\
 & $0.03, 0.03, 0.02, 0.02, 0.02, 0.01, 0.01, 0.01, 0.01, 0.01)$ \\
Dominant 0.20 & $(0.20, 0.12, 0.10, 0.08, 0.07, 0.06, 0.05, 0.04, 0.03, 0.03,$ \\
 & $0.03, 0.02, 0.02, 0.02, 0.01, 0.01, 0.01, 0.01, 0.01, 0.01)$ \\
\hline
\end{tabular}
\label{tab:bids_20}
\end{table}

\end{document}